\documentclass[aps,floatfix,superscriptaddress,preprint]
{revtex4}

\usepackage{titlesec}
\usepackage{amssymb,amsmath,amsfonts,latexsym,graphicx,epsfig,bm}
\usepackage{epstopdf}

\titleformat{\section}{\large\bfseries}{\thesection}{1em}{}

\newcommand{\bea}{\begin{eqnarray}}
\newcommand{\ena}{\end{eqnarray}}
\newcommand{\nn}{\nonumber\\}
\newcommand{\be}{\begin{equation}}
\newcommand{\en}{\end{equation}}

\newcommand{\ed}{\end{document}}

\newcommand{\slp}{p\kern-5pt/}
\newcommand{\Tr}{\mbox{\rm{tr}}}

\begin{document}

\title{Implications of new physics in the decays 
\boldmath{$B_c \to (J/\psi,\eta_c)\tau\nu$}}

\author{C. T. Tran}
\email{ctt@theor.jinr.ru,tranchienthang1347@gmail.com}
\affiliation{Institute of Research and Development, Duy Tan University, 550000 Da Nang, Vietnam}
\affiliation{
Bogoliubov Laboratory of Theoretical Physics,
Joint Institute for Nuclear Research,
141980 Dubna, Russia}

\author {M. A. Ivanov}
\email{ivanovm@theor.jinr.ru}
\affiliation{
Bogoliubov Laboratory of Theoretical Physics,
Joint Institute for Nuclear Research,
141980 Dubna, Russia}

\author{J. G. K\"{o}rner}
\email{jukoerne@uni-mainz.de}
\affiliation{PRISMA Cluster of Excellence, Institut f\"{u}r Physik, 
Johannes Gutenberg-Universit\"{a}t, 
D-55099 Mainz, Germany}

\author{P. Santorelli}
\email{Pietro.Santorelli@na.infn.it}
\affiliation{Dipartimento di Fisica, Universit\`{a} di Napoli Federico II, Complesso Universitario di Monte S. Angelo, Via Cintia, Edificio 6, 80126 Napoli, Italy}
\affiliation{Istituto Nazionale di Fisica Nucleare, Sezione di Napoli, 80126 Napoli, Italy}

\date{\today}

\begin{abstract}
We study the semileptonic decays of the $B_c$  meson into final charmonium states within the standard model and beyond. The relevant hadronic transition form factors are calculated in the framework of the covariant confined quark model developed by us. We focus on the tau mode of these decays, which may provide some hints of new physics effects. We extend the standard model by assuming a general effective Hamiltonian describing the $b\to c\tau\nu$ transition, which consists of the full set of the four-fermion operators. We then obtain experimental constraints on the Wilson coefficients corresponding to each operator and provide predictions for the branching fractions and other polarization observables in different new physics scenarios.
\end{abstract}



\maketitle

\section{Introduction}
\label{sec:intro} 
The $B_c$ meson is the lowest bound state of two heavy quarks of different flavors, lying below the $B\bar D$ threshold. As a result, while the corresponding $c\bar c$ and $b\bar b$ quarkonia decay strongly and electromagnetically, the $B_c$ meson decays weakly, making it possible to study weak decays of doubly heavy mesons. The weak decays of the $B_c$ meson proceed via the $c$-quark decays $(\sim 70\%)$, the $b$-quark decays $(\sim 20\%)$, and the weak annihilation $(\sim 10\%)$. Due to its outstanding features, the $B_c$ meson and its decays have been studied extensively (for a review, see e.g. Ref.~\cite{Gouz:2002kk} and references therein).

Among many weak decays of the $B_c$ meson, the semileptonic decay $B_c \to J/\psi \ell\nu$ has an important meaning. In fact, the first observation of the $B_c$ meson by the CDF Collaboration was made in an analysis of this decay~\cite{Abe:1998fb}. Recently, the LHCb Collaboration has reported their measurement~\cite{Aaij:2017tyk} of the ratio of branching fractions
\be
R_{J/\psi}\equiv \frac{\mathcal{B}(B_c \to J/\psi \tau\nu)}{\mathcal{B}(B_c \to J/\psi \mu\nu)}=0.71\pm 0.17\pm 0.18,
\en
which lies at about $2\sigma$ above the range of existing predictions in the Standard Model (SM). At the quark level, the decay $B_c \to J/\psi \ell\nu$ is described by the transition $b\to c\ell\nu$, which is identical to that of the decays $\bar{B}^0 \to D^{(\ast)} \ell\nu$. It is important to note that measurements of the decays $\bar{B}^0 \to D^{(\ast)} \ell\nu$ carried out by the {\it BABAR}~\cite{Lees:2013uzd}, Belle~\cite{Huschle:2015rga,Sato:2016svk,Hirose:2016wfn}, and LHCb~\cite{Aaij:2015yra, Aaij:2017uff} Collaborations have also revealed a significant deviation ($\sim 4\sigma$) of the ratios $R_{D^{(\ast)}}$ from the SM predictions~\cite{Na:2015kha, Lattice:2015rga, Fajfer:2012vx, Amhis:2016xyh}.
The excess of $R_{J/\psi}$ over the SM predictions not only sheds more light on the unsolved $R_{D^{(\ast)}}$ puzzle, but also suggests the consideration of possible new physics (NP) effects in the decays $B_c \to J/\psi(\eta_c) \tau\nu$.

Essential to the study of the $B_c$ semileptonic decays is the calculation of the invariant form factors describing the corresponding hadronic transitions. In the literature, a wide range of different approaches has been used to compute the $B_c\to J/\psi (\eta_c)$ transition form factors, such as the potential model approach~\cite{Chang:1992pt}, the Bethe-Salpeter equation~\cite{AbdElHady:1999xh, Liu:1997hr}, the relativistic constituent quark model on the light front~\cite{Anisimov:1998xv, Wang:2008xt}, three-point sum rules of QCD and nonrelativistic QCD~\cite{Kiselev:1999sc, Kiselev:2000pp,Kiselev:2002vz}, the relativistic quark model based on the quasipotential approach~\cite{Ebert:2003cn}, the nonrelativistic quark model~\cite{Hernandez:2006gt}, the Bauer-Stech-Wirbel framework~\cite{Dhir:2008hh}, the perturbative QCD (pQCD)~\cite{Wen-Fei:2013uea, Rui:2016opu}, and the covariant quark model developed by our group~\cite{Ivanov:2000aj, Ivanov:2005fd}. It is worth revisiting these decays in the modern version of our model with updated parameters and new features like the embedded infrared 
confinement~\cite{Issadykov:2017wlb}. We also mention that very recently, the HPQCD Collaboration has provided their preliminary results for the form factors of the $B_c\to J/\psi$ and $B_c\to \eta_c$ transitions using lattice QCD~\cite{Colquhoun:2016osw}. 

It should be noted that in our covariant confined quark model (CCQM), all form factors are calculated in the full kinematic range of momentum transfer squared $q^2$, making the predictions for physical observables more accurate. In the pQCD approach and QCD sum rules, for example, the form factors are evaluated only for small values of $q^2$ (large recoil), and then extrapolated to the large $q^2$ region (small recoil), in which they become less reliable. In general, the knowledge of the $B_c\to J/\psi (\eta_c)$ form factors is much less than that of the $\bar{B}^0\to D^{(\ast)}$ ones. This is due to, first, the lack of experimental data for the decays $B_c\to J/\psi (\eta_c)\ell\nu$, and second, the appearance of two heavy quark flavors in the initial ($b\bar{c}$) and final ($c\bar{c}$) states. The latter breaks the heavy quark symmetry (HQS), leaving the residual heavy quark spin symmetry (HQSS), which allows reducing the number of form factors in the infinite heavy quark limit~\cite{Jenkins:1992nb,Colangelo:1999zn,Kiselev:1999sc}. However, the HQSS does not fix the normalization of the form factors as the HQS does, for example, in the case of $\bar{B}^0\to D^{(\ast)}$.

The possible NP effects in the semileptonic decays $B_c \to J/\psi(\eta_c) \tau\nu$ have been discussed recently in several papers~\cite{Dutta:2017xmj, Watanabe:2017mip, Chauhan:2017uil, Dutta:2017wpq, Alok:2017qsi, He:2017bft, Wei:2018vmk,Biswas:2018jun}. As what has been done in the studies of the $R_{D^{(\ast)}}$ anomalies, one can choose between a specific-model approach, such as charged Higgs models, leptoquark models etc., or a model-independent approach based on a general effective Hamiltonian describing the $b\to c\tau\nu$ transition. In this paper we adopt the second approach by proposing an SM-extended effective Hamiltonian consisting of the full set of the four-fermion operators. Constraints on the corresponding Wilson coefficients are obtained from the experimental data for the ratios $R_{J/\psi}$ and $R_{D^{(\ast)}}$, as well as the LEP1 data, which requires $\mathcal{B}(B_c\to\tau\nu)\leq 10\%$~\cite{Akeroyd:2017mhr}. Another useful constraint is provided by using the lifetime of $B_c$ as discussed in Ref.~\cite{Alonso:2016oyd}. However, in this paper we only use the constraint from the LEP1 data, which is more stringent than the latter. We then analyze the effects of these NP operators on several physical observables including the ratios of branching fractions $R_{J/\psi(\eta_c)}$, the forward-backward asymmetries, the convexity parameter, and the polarization components of the $\tau$ in the final state. We also provide our predictions for these physical observables in the SM and in the presence of NP. 

The paper is organized as follows. In Sec.~\ref{sec:distr} we introduce the general $b\to c\ell\nu$ effective Hamiltonian and parametrize the hadronic matrix elements in terms of the invariant form factors. We then  obtain the decay distributions in the presence of NP operators using the helicity technique. In Sec.~\ref{sec:FF} we present our result for the form factors in the whole $q^2$ range. A detailed comparison of the form factors calculated in the CCQM with those in other approaches is also provided. In Sec.~\ref{sec:constraint} we obtain constraints on the NP Wilson coefficients from available experimental data. Theoretical predictions for the physical observables in the SM and beyond are presented in Sec.~\ref{sec:prediction}. Finally, we briefly conclude in Sec.~\ref{sec:summary}.

\section{Effective Hamiltonian, helicity amplitudes, and decay distribution}
\label{sec:distr}
In the model-independent approach, NP effects are introduced explicitly by proposing an effective Hamiltonian for the weak decays that  includes both SM and beyond-SM contributions. In this study, the general effective Hamiltonian for the quark-level
transition $b \to c \ell \nu$ $(\ell=e,\mu,\tau)$ is given by $(i=L,R)$
\bea
\label{eq:Heff}
\mathcal{H}_{eff} &=&\frac{4G_F V_{cb}}{\sqrt{2}} \Big(\mathcal{O}_{V_L}+
\sum\limits_{X=S_i,V_i,T_L} \delta_{\tau\ell}X\mathcal{O}_{X}\Big),
\ena
where the four-fermion operators $\mathcal{O}_{X}$ are defined as 
\bea
\mathcal{O}_{V_i} &=&
\left(\bar{c}\gamma^{\mu}P_ib\right)
\left(\bar{\ell}\gamma_{\mu}P_L\nu_{\ell}\right),\\
\mathcal{O}_{S_i} &=& \left(\bar{c}P_ib\right)\left(\bar{\ell}P_L\nu_{\ell}\right),
\\
\mathcal{O}_{T_L} &=& \left(\bar{c}\sigma^{\mu\nu}P_Lb\right)
\left(\bar{\ell}\sigma_{\mu\nu}P_L\nu_{\ell}\right).
\ena
Here, $\sigma_{\mu\nu}=i\left[\gamma_{\mu},\gamma_{\nu}\right]/2$, 
$P_{L,R}=(1\mp\gamma_5)/2$ are the left and right projection operators, and $X$'s are the complex Wilson coefficients characterizing the NP contributions. The tensor operator with right-handed quark current is identically equal to zero and is therefore omitted. In the SM one has $V_{L,R}=S_{L,R}=T_L=0$. We have assumed that neutrinos are left-handed. Besides, the delta function in Eq.~(\ref{eq:Heff}) implies that NP effects are supposed to appear in the tau mode only. The proposed Hamiltonian can be considered as a natural way to go beyond the SM since it is generalized from the well established SM Hamiltonian with the $V-A$ structure by adding more currents. One may also consider right-handed neutrinos and may as well assume that NP appears in all lepton generations. However, current  experimental data suggest that NP effects in the case of light leptons (if any) are very small. A recent discussion of these NP operators and their possible appearance in the light lepton modes can be found in Ref.~\cite{Jung:2018lfu}.

Starting with the effective Hamiltonian, one writes down the matrix element of the semileptonic decays $B_c \to J/\psi(\eta_c) \tau\nu$, which has the form
\be
\mathcal{M}=\mathcal{M}_{\rm SM}+\sqrt{2} G_F V_{cb}\sum\limits_{X}
X\cdot
\langle J/\psi(\eta_c)
|\bar{c} \Gamma_X b
|B_c \rangle
\cdot
\bar\tau \Gamma_X \nu_\tau,
\en
where $\Gamma_X$ is the Dirac matrix corresponding to the operator $\mathcal{O}_X$. The hadronic part in the matrix element is parametrized by a set of invariant form factors depending on the momentum transfer squared $q^2$ between the two hadrons as follows:
\bea
\langle \eta_c(p_2)
|\bar{c} \gamma^\mu b
| B_c(p_1) \rangle
&=& F_+(q^2) P^\mu + F_-(q^2) q^\mu,\nn
\langle \eta_c(p_2)
|\bar{c}b
| B_c(p_1) \rangle &=& (m_1+m_2)F^S(q^2),\nn
\langle \eta_c(p_2)|\bar{c}\sigma^{\mu\nu}(1-\gamma^5)b|B_c(p_1)\rangle 
&=&\frac{iF^T(q^2)}{m_1+m_2}\left(P^\mu q^\nu - P^\nu q^\mu 
+i \varepsilon^{\mu\nu Pq}\right),\nn
\langle J/\psi(p_2)
|\bar{c} \gamma^\mu(1\mp\gamma^5)b
| B_c(p_1) \rangle
&=& \frac{\epsilon^{\dagger}_{2\alpha}}{m_1+m_2}
\Big[ \mp g^{\mu\alpha}PqA_0(q^2) \pm P^{\mu}P^{\alpha}A_+(q^2)\nn
&&\pm q^{\mu}P^\alpha A_-(q^2) 
+ i\varepsilon^{\mu\alpha P q}V(q^2)\Big],
\nn
\langle J/\psi(p_2)
|\bar{c}\gamma^5 b
| B_c(p_1) \rangle &=& \epsilon^\dagger_{2\alpha}P^\alpha G^P(q^2),
\nn
\langle J/\psi(p_2)|\bar{c}\sigma^{\mu\nu}(1-\gamma^5)b|B_c(p_1)\rangle
&=&-i\epsilon^\dagger_{2\alpha}\Big[
\left(P^\mu g^{\nu\alpha} - P^\nu g^{\mu\alpha} 
+i \varepsilon^{P\mu\nu\alpha}\right)G_1^T(q^2)\nn
&&+\left(q^\mu g^{\nu\alpha} - q^\nu g^{\mu\alpha}
+i \varepsilon^{q\mu\nu\alpha}\right)G_2^T(q^2)\nn
&&+\left(P^\mu q^\nu - P^\nu q^\mu 
+ i\varepsilon^{Pq\mu\nu}\right)P^\alpha\frac{G_0^T(q^2)}{(m_1+m_2)^2}
\Big],
\label{eq:ff}
\ena
where $P=p_1+p_2$, $q=p_1-p_2$, and $\epsilon_2$ is the polarization vector
of the $J/\psi$ meson which satisfies the condition $\epsilon_2^\dagger\cdot p_2=0$.
The particles are on their mass shells: $p_1^2=m_1^2=m_{B_c}^2$ and
 $p_2^2=m_2^2=m_{J/\psi(\eta_c)}^2$.

We define a polar angle $\theta$ as the angle between ${\bf q}={\bf p_1}-{\bf p_2}$ and the three-momentum of the charged lepton in the ($\ell\bar\nu_\ell$) rest frame. The angular decay distribution then reads
\be
\frac{d\Gamma}{dq^2 d\cos\theta} = 
\frac{|{\bf p_2}|}{(2\pi)^3 32 m_1^2} \Big(1-\frac{m^2_\ell}{q^2}\Big)
\sum\limits_{\rm{pol}}|{\cal M}|^2
=\frac{G^2_F |V_{cb}|^2 |{\bf p_2}|}{(2\pi)^3 64 m_1^2}\Big(1-\frac{m^2_\ell}{q^2}\Big)
H^{\mu\nu} L_{\mu\nu},
\label{eq:2-fold-dis}
\en
where  $|{\bf p_2}|=\lambda^{1/2}(m_1^2,m_2^2,q^2)/2m_1 $
is the momentum of the daughter meson in the $B_c$ rest frame, and $H^{\mu\nu}L_{\mu\nu}$ is the contraction 
of hadron and lepton tensors. 
The covariant contraction $H^{\mu\nu} L_{\mu\nu}$ can be converted to a sum
of bilinear products of hadronic and leptonic helicity amplitudes using the 
completeness relation for the polarization four-vectors of the process~\cite{Korner:1989qb}. This technique is known as the helicity technique, which has been described in great detail in our previous papers~\cite{Korner:1989ve,Korner:1989qb,Ivanov:2015tru,Ivanov:2016qtw}. In 
Ref.~\cite{Ivanov:2016qtw} we have shown how to acquire the decay distribution for the semileptonic decays $\bar{B}^0\to D^{(\ast)}\tau\nu$ in the presence of NP operators and provided a full description of the helicity amplitudes, which can be applied to the case of the decays $B_c\to J/\psi(\eta_c)\tau\nu$. Therefore, we find no reason to repeat the procedure in this paper. However, for completeness, we present here the final result for the decay distributions. The angular distribution for the decay $B_c\to \eta_c\tau\nu$ is written as follows:
\bea
\lefteqn{\frac{d\Gamma(B_c\to \eta_c\tau\nu)}{dq^2d\cos\theta}}\nn
&=&\frac{G_F^2|V_{cb}|^2|{\bf p_2}|q^2}{(2\pi)^3 16m_1^2}\Big(1-\frac{m^2_\tau}{q^2}\Big)^2 \nn
&&\times 
\big\lbrace|1+g_V|^2\left[ |H_0|^2\sin^2\theta +2\delta_\tau |H_t-H_0\cos\theta|^2 \right]\nn
&&+|g_S|^2|H_P^S|^2
+16|T_L|^2\left[2\delta_\tau+( 1-2\delta_\tau)\cos^2\theta \right]|H_T|^2\nn
&&+2\sqrt{2\delta_\tau} {\rm Re}g_S H_P^S\left[ H_t-H_0\cos\theta \right]
+8\sqrt{2\delta_\tau} {\rm Re}T_L\left[ H_0-H_t\cos\theta \right]H_T
\big\rbrace,
\label{eq:distr2D}
\ena
where $g_V\equiv V_L+V_R$, $g_S\equiv S_L+S_R$, $g_P\equiv S_L-S_R$, and $\delta_\tau = m^2_\tau/2q^2$ is the helicity flip factor.
The hadronic helicity amplitudes $H$'s are written in terms of the invariant form factors defined in Eq.~(\ref{eq:ff}). Their explicit expressions are presented in Ref.~\cite{Ivanov:2016qtw}. Note that we do not consider interference terms between different NP operators since we assume the dominance of only one NP operator besides the SM contribution. 
The corresponding distribution for the decay $B_c\to J/\psi\tau\nu$ is rather cumbersome and therefore is not shown here. One can find it in  Appendix~C of Ref.~\cite{Ivanov:2016qtw}. 

After integrating the angular distribution over $\cos\theta$ one has 
\be
\frac{d\Gamma(B_c\to J/\psi(\eta_c)\tau\nu)}{dq^2}
=\frac{G_F^2|V_{cb}|^2|{\bf p_2}|q^2}{(2\pi)^3 12m_1^2}\Big(1-\frac{m^2_\tau}{q^2}\Big)^2\cdot {\cal H}_{tot}^{J/\psi(\eta_c)},\quad \text{where}
\label{eq:distr1}
\en
\bea
{\cal H}_{tot}^{\eta_c}
&=&
|1+g_V|^2\left[|H_0|^2+\delta_\tau(|H_0|^2+3|H_t|^2) \right]+\frac{3}{2}|g_S|^2 |H_P^S|^2\nn
&&+ 3\sqrt{2\delta_\tau} {\rm Re}g_S H_P^S H_t
+8|T_L|^2 ( 1+4\delta_\tau) |H_T|^2
+12\sqrt{2\delta_\tau} {\rm Re}T_L H_0 H_T,\\[1.0em]
{\cal H}_{tot}^{J/\psi}
&=&(|1+V_L|^2+|V_R|^2)\left[\sum\limits_{n=0,\pm}|H_{n}|^2+\delta_\tau \left(\sum\limits_{n=0,\pm}|H_{n}|^2+3|H_{t}|^2\right) \right]+\frac{3}{2}|g_P|^2|H^S_V|^2\nn
&&-2 {\rm Re}V_R\big[(1+\delta_\tau) (|H_{0}|^2+2H_{+}H_{-})+3\delta_\tau |H_{t}|^2 \big]
-3\sqrt{2\delta_\tau} {\rm Re}g_P H^S_V H_{t}\nn
&&+8|T_L|^2 (1+4\delta_\tau)\sum\limits_{n=0,\pm}|H_T^n|^2
-12\sqrt{2\delta_\tau} {\rm Re}T_L\sum\limits_{n=0,\pm} H_{n}H_T^n.
\ena

In this paper, we also impose the constraint from the leptonic decay channel of $B_c$ on the Wilson coefficients. Therefore we present here the leptonic branching in the presence of NP operators. In the SM, the purely leptonic decays $B_c\to \ell \nu$ 
proceed via the annihilation of the quark pair into an off shell $W$ boson. Assuming the effective Hamiltonian Eq.~(\ref{eq:Heff}), the tau mode of these decays receives NP contributions from all operators except $\mathcal{O}_{T_L}$. The branching fraction of the leptonic decay in the presence of NP is given by~\cite{Ivanov:2017hun}
\be
 \mathcal{B}(B_c \to \tau \nu)=
\frac{G_F^2}{8\pi}|V_{cb}|^2\tau_{B_c}m_{B_c}m_{\tau}^2\left(1-\frac{m_{\tau}^2}{m_{B_c}^2}\right)^2f_{B_c}^2
 \times
\left|
1-g_A+\frac{m_{B_c}}{m_\tau} \frac{f_{B_c}^P}{f_{B_c}}g_P
\right|^2,
\en
where $g_A\equiv V_R-V_L$, $g_P\equiv S_R-S_L$, $\tau_{B_c}$ is the $B_c$ lifetime, $f_{B_c}$ is the leptonic decay constant of $B_c$, and $f_{B_c}^P$ is a new  constant corresponding to the new quark current structure. One has
\be
\langle 0
|\bar{q} \gamma^\mu \gamma_5 b
| B_c(p) \rangle
= -f_{B_c} p^\mu,\qquad
\langle 0
|\bar{q}\gamma_5b
| B_c(p) \rangle = m_{B_c} f_{B_c}^P.
\en
In the CCQM, we obtain the following values for these constants (all in MeV):
\be
f_{B_c}=489.3,\quad f_{B_c}^P=645.4.
\en

\section{Form factors in the covariant confined quark model}
\label{sec:FF}

The CCQM is an effective quantum field approach to hadron physics, which is based on a relativistic invariant Lagrangian describing the interaction of a hadron with its constituent quarks (see e.g. Refs.~\cite{Efimov:1988yd, Efimov:zg, Branz:2009cd, Ivanov:2011aa,Gutsche:2012ze,Gutsche:2013oea,Ivanov:2015woa}). The hadron is described by a field  $H(x)$, which satisfies the corresponding equation of motion, while the quark part is introduced by an interpolating quark current $J_H(x)$ with the hadron quantum numbers. In the case of mesons, the Lagrangian is written as
\be
{\mathcal L}_{\rm int}(x) = g_H H(x) J_H(x)
=g_H H(x)\int\!\! dx_1 \!\!\int\!\! dx_2 F_H (x;x_1,x_2)
\bar q_2(x_2)\Gamma_H q_1(x_1),
\label{eq:Lagr}
\en
where $g_H$ is the quark-meson coupling, $\Gamma_H$ is the Dirac matrix ensuring the quantum numbers of the meson, and the so-called vertex function $F_H$ effectively describes the quark distribution inside the meson. From the requirement for the translational invariance of $F_H$, we adopt the following form:
$
F_H(x,x_1,x_2)=\delta(x - w_1 x_1 - w_2 x_2) \Phi_H((x_1-x_2)^2),
$
where $w_i = m_{q_i}/(m_{q_1}+m_{q_2})$, and $m_{q_i}$ are the constituent quark masses. The Fourier transform of the function $\Phi_H$ in momentum space is required to fall off in the Euclidean region in order to provide for the ultraviolet convergence of the loop integrals. For the sake of simplicity we use the Gaussian form 
$\widetilde\Phi_H(-k^2)=\exp(k^2/\Lambda_H^2)$, where the parameter $\Lambda_H$ 
effectively characterizes the meson size.

\begin{figure}[htbp]
\begin{center}
\includegraphics[scale=0.5]{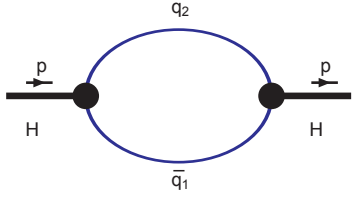}
\caption{\small Self-energy diagram for a meson.}
\label{fig:mass}
\end{center}
\end{figure}

The coupling $g_H$ is determined by using the so-called compositeness condition~\cite{Z=0}, which imposes that the wave function renormalization constant of the hadron is equal to zero $Z_H=0$. For mesons, the condition has the form
$Z_H=1-\Pi'_H (m^2_H)=0$, where $\Pi'_H (m^2_H)$ is the derivative of the hadron mass operator, which corresponds to the self-energy diagram in Fig.~\ref{fig:mass}
and has the following form
\bea
\Pi_P(p^2) &=& 3g_P^2 \int\!\! \frac{dk}{(2\pi)^4i}
\widetilde\Phi^2_P \left(-k^2\right)
\Tr\left[ S_1(k+w_1p)\gamma^5 S_2(k-w_2p)\gamma^5 \right],
\nn
\Pi_V(p^2) &=& g_V^2 \Big(g^{\mu\nu} - \frac{p^{\mu}p^{\nu}}{p^2}\Big) 
\int\!\! \frac{dk}{(2\pi)^4i}\widetilde\Phi^2_V \left(-k^2\right)
\Tr\left[ S_1(k+w_1p)\gamma_{\mu} S_2(k-w_2p)\gamma_{\nu} \right],
\ena
for pseudoscalar and vector mesons, respectively. Here, $S_{1,2}$ are quark propagators, for which we use the Fock-Schwinger representation
\be
S_i(k) = (m_{q_i}+\not\! k)
\int\limits_0^\infty\! d\alpha_i \exp[-\alpha_i(m_{q_i}^2-k^2)].
\en
It should be noted that all loop integrations  are carried out in Euclidean space.

Similarly to the hadron mass operator, matrix elements of hadronic transitions are represented by quark-loop diagrams, which are described as convolutions of the corresponding quark propagators and vertex functions. Using various techniques described in our previous papers, any hadronic matrix element $\Pi$ can be finally written in the form
$\Pi =  \int\limits_0^\infty\! d^n \alpha\, F(\alpha_1,\ldots,\alpha_n)$, where $F$
is the resulting integrand corresponding to a given diagram. It is more convenient to turn the set of Fock-Schwinger parameters into a simplex by adding the integral
$1=\int\limits_0^\infty\!dt\,\delta\big(t-\sum\limits_{i = 1}^2 {{\alpha _i}}\big)$
as follows:
\be
\Pi   = \int\limits_0^\infty\! dt t^{n-1} \int\limits_0^1\! d^n \alpha  
\delta\Big(1 -  \sum\limits_{i=1}^n \alpha_i \Big)  
F(t\alpha_1,\ldots,t\alpha_n). 
\label{eq:Pi}
\en
The integral in Eq.~(\ref{eq:Pi}) begins to diverge when $t\to\infty$, if the kinematic variables allow the appearance of branching point corresponding to the creation of free quarks. However, these possible threshold singularities disappear if one cuts off the integral at the upper limit:
\be
\Pi^{\rm c}   = \int\limits_0^{1/\lambda^2}\! dt t^{n-1} 
\int\limits_0^1\! d^n \alpha  
\delta\Big(1 -  \sum\limits_{i=1}^n \alpha_i \Big) 
F(t\alpha_1,\ldots,t\alpha_n). 
\en
The parameter $\lambda$ effectively guarantees the confinement of quarks inside a hadron and is called the infrared cutoff parameter.

Finally, we briefly discuss some error estimates within our model.
The CCQM consists of several free parameters: the constituent quark masses $m_q$, the hadron size parameters $\Lambda_H$, and the universal infrared cutoff parameter $\lambda$. These parameters are determined by
minimizing the functional
$\chi^2 = \sum\limits_i\frac{(y_i^{\rm expt}-y_i^{\rm theor})^2}{\sigma^2_i}$
where $\sigma_i$ is the experimental  uncertainty.
If $\sigma$ is too small then we take its value of 10$\%$.
Besides, we have observed that the errors of the fitted parameters 
are of the order of  10$\%$.
Thus, the theoretical error of the CCQM is estimated to be of the order of 10$\%$.

\begin{figure}[htbp]
\begin{tabular}{lr}
\includegraphics[scale=0.5]{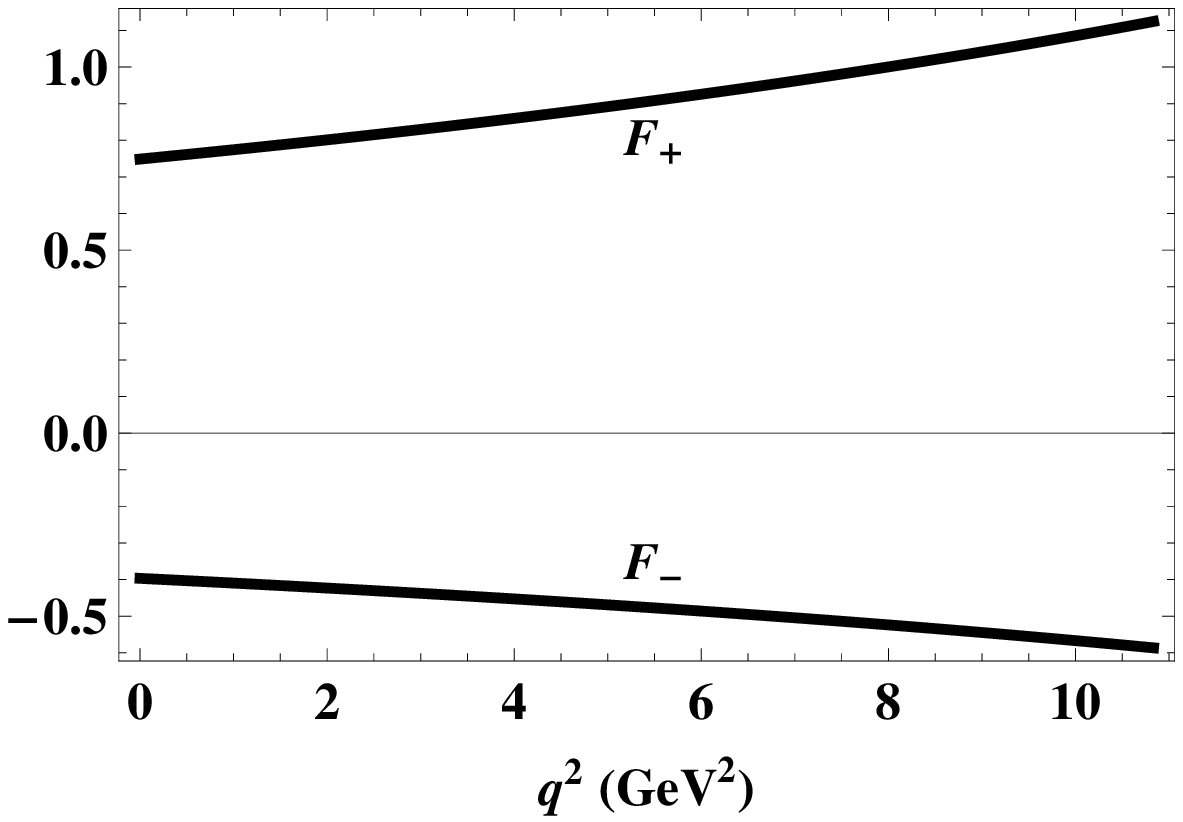}
& 
\includegraphics[scale=0.5]{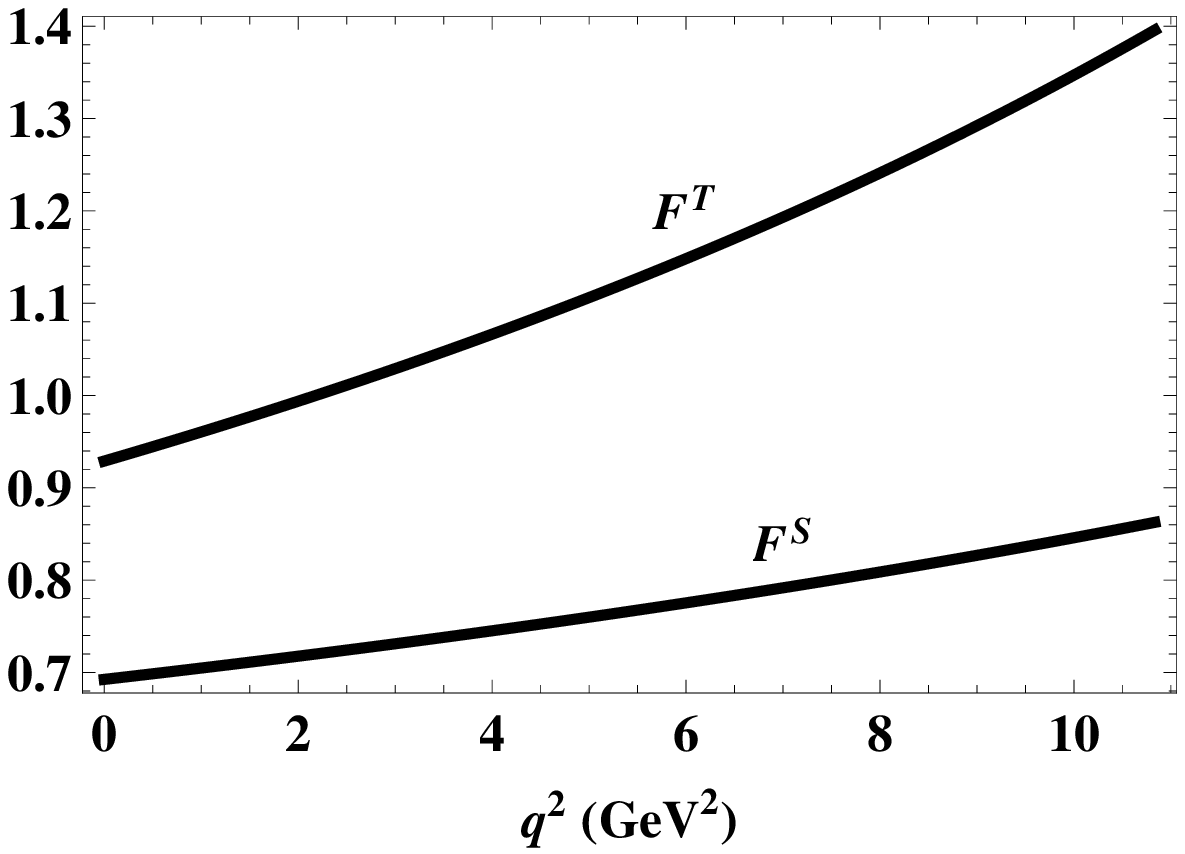}\\
\includegraphics[scale=0.5]{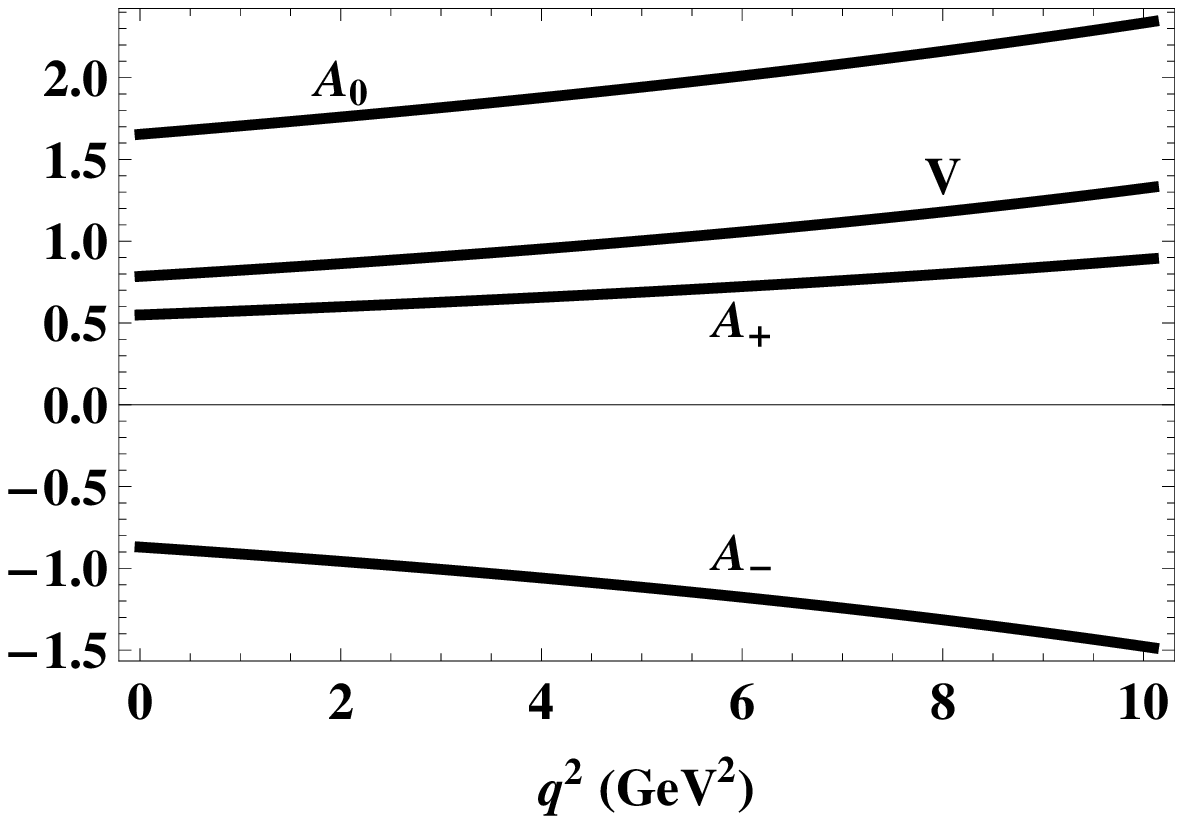}
& 
\includegraphics[scale=0.5]{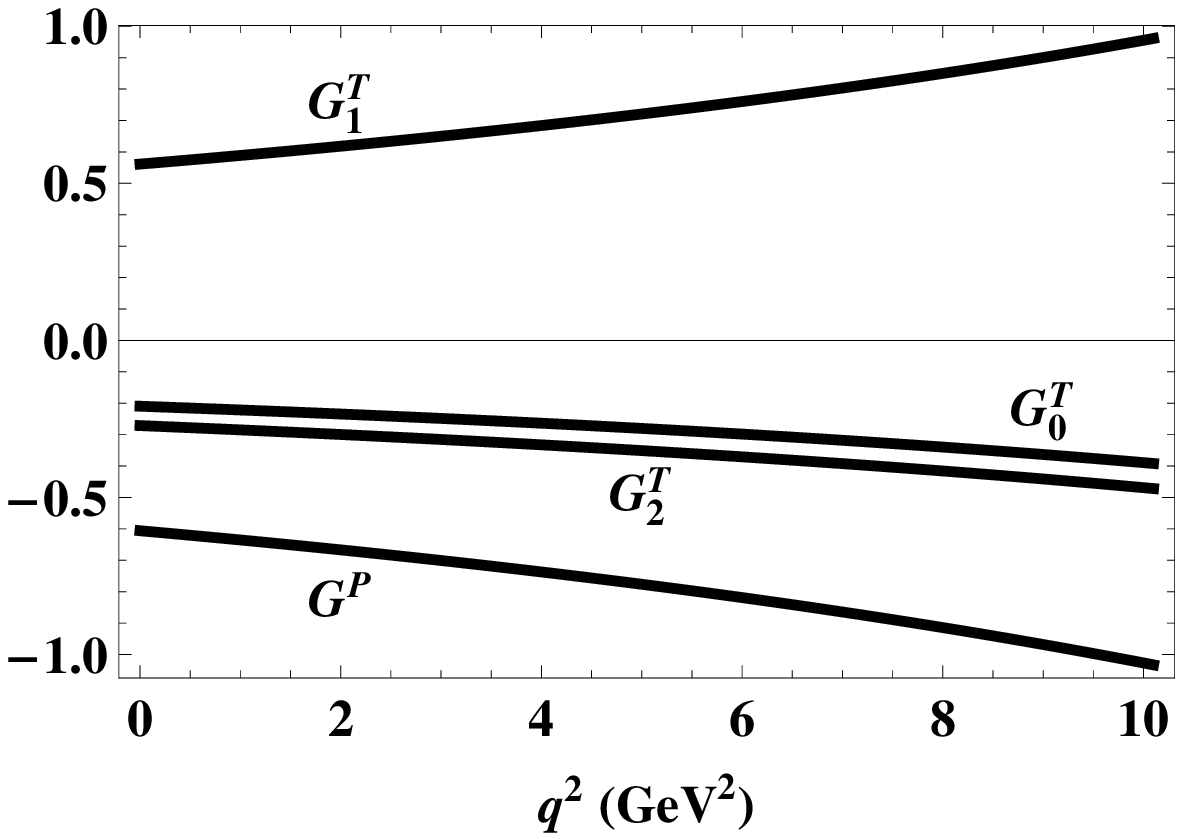}
\end{tabular}
\caption{Form factors of the transitions $B_c\to\eta_c$ (upper panels)
  and $B_c \to J/\psi$ (lower panels).}
\label{fig:formfactor}
\end{figure}

The $B_c\to J/\psi(\eta_c)$ hadronic transitions are calculated from their one-loop quark diagrams. For a more detailed description of the calculation techniques we refer to Ref.~\cite{Ivanov:2016qtw} where we computed the similar form factors for the $\bar{B}^0\to D^{(\ast)}$ transitions. In the framework of the CCQM, the interested form factors  are represented by threefold integrals
which are calculated by using \textsc{fortran} codes in the full kinematical momentum 
transfer region $0\le q^2 \le q^2_{max}=(m_{B_c}-m_{J/\psi(\eta_c)})^2$. The numerical results for the form factors are well approximated by a double-pole parametrization
\be
F(q^2)=\frac{F(0)}{1 - a s + b s^2}, \quad s=\frac{q^2}{m_{B_c}^2}. 
\label{eq:ff-para}
\en
The parameters of the $B_c \to J/\psi(\eta_c)$ form factors  are listed in Table~\ref{tab:ff-param}. Their $q^2$ dependence in the full momentum transfer range $0\le q^2 \le q^2_{max}=(m_{B_c}-m_{J/\psi(\eta_c)})^2$ is shown in Fig.~\ref{fig:formfactor}.

Firstly, we focus on those form factors that are needed to describe the $B_c \to J/\psi(\eta_c)$ transitions within the SM (without any NP operators), namely, $F_{\pm}$, $A_{0,\pm}$, and $V$. It is worth noting that all these form factors have a pronounced $(q^2)^{-2}$ contribution (the ratio $b/a$ lies between $0.14$ and $0.50$) in comparison with the case $\bar{B}^0\to D^{(\ast)}$, where all form factors (except for $A_0$) have a very small ratio $b/a\sim 0.05{-}0.08$, and therefore show a monopolelike behavior~\cite{Ivanov:2015tru}.
\begin{table}[htbp]
\caption{Parameters of the dipole approximation in Eq.~(\ref{eq:ff-para}) for  $B_c \to J/\psi(\eta_c)$ form factors. Zero-recoil (or $q^2_{\rm max}$) values of the form factors are also listed.}
\begin{center}
\begin{tabular}{lccccccccccccc}
\hline\hline
\multicolumn{1}{c}{} &\multicolumn{8}{c}{$B_c \to J/\psi$} &\multicolumn{1}{c}{} 
                      &\multicolumn{4}{c}{$B_c \to \eta_c$} \\
\cline{2-9}\cline{11-14}
 & $ A_0 $ & $  A_+  $ & $  A_-  $ & $  V  $ 
 & $ G^P $ & $G_0^T$ & $G_1^T$ & $  G_2^T$ & {} & $F_+$ & $F_-$ & $F^S$ & $F^T$ 
 \\
\hline
$F(0)$ &  1.65 & 0.55  & $-0.87$ & 0.78 & $-0.61$ & $-0.21$ & 0.56 & $-0.27$ & {} &  0.75   & $-0.40$ &  0.69 & 0.93  
\\
$a$    &  1.19 & 1.68  &  1.85 & 1.82 & 1.84 & 2.16 & 1.86 & 1.91 & {} &  1.31   &  1.25 &  0.68 & 1.30  
\\
$b$    & 0.17 & 0.70 & 0.91 & 0.86 & 0.91 & 1.33 & 0.93 & 1.00 & {} &  0.33  & 0.25 & $-0.12$ & 0.31 
\\ 
$F(q^2_{\rm max})$ &  2.34 & 0.89  & $-1.49$ & 1.33 & $-1.03$ & $-0.39$ & 0.96 & $-0.47$ & {} &  1.12   & $-0.59$ &  0.86 & 1.40  
\\
\hline\hline
\end{tabular}
\label{tab:ff-param}
\end{center}
\end{table}

\begin{figure}[htbp]
\begin{tabular}{lr}
\includegraphics[scale=0.6]{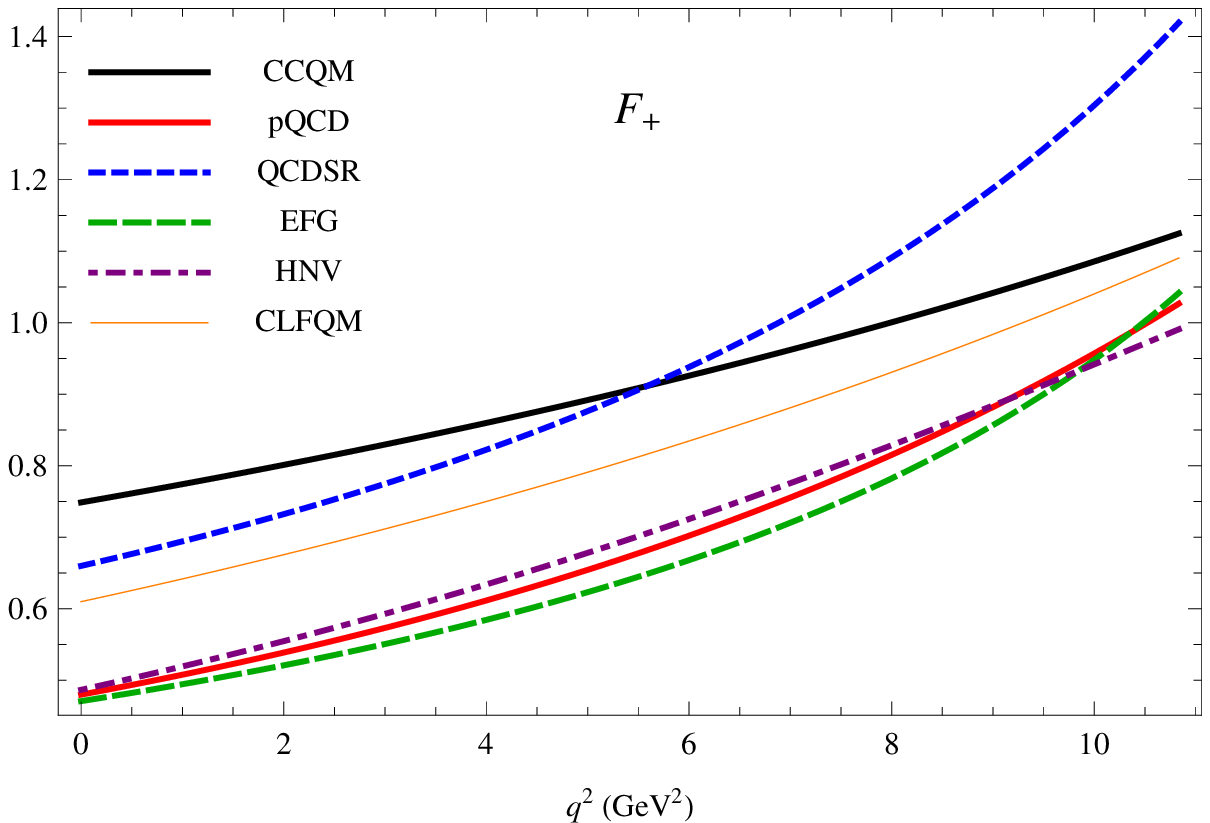}
&
\includegraphics[scale=0.6]{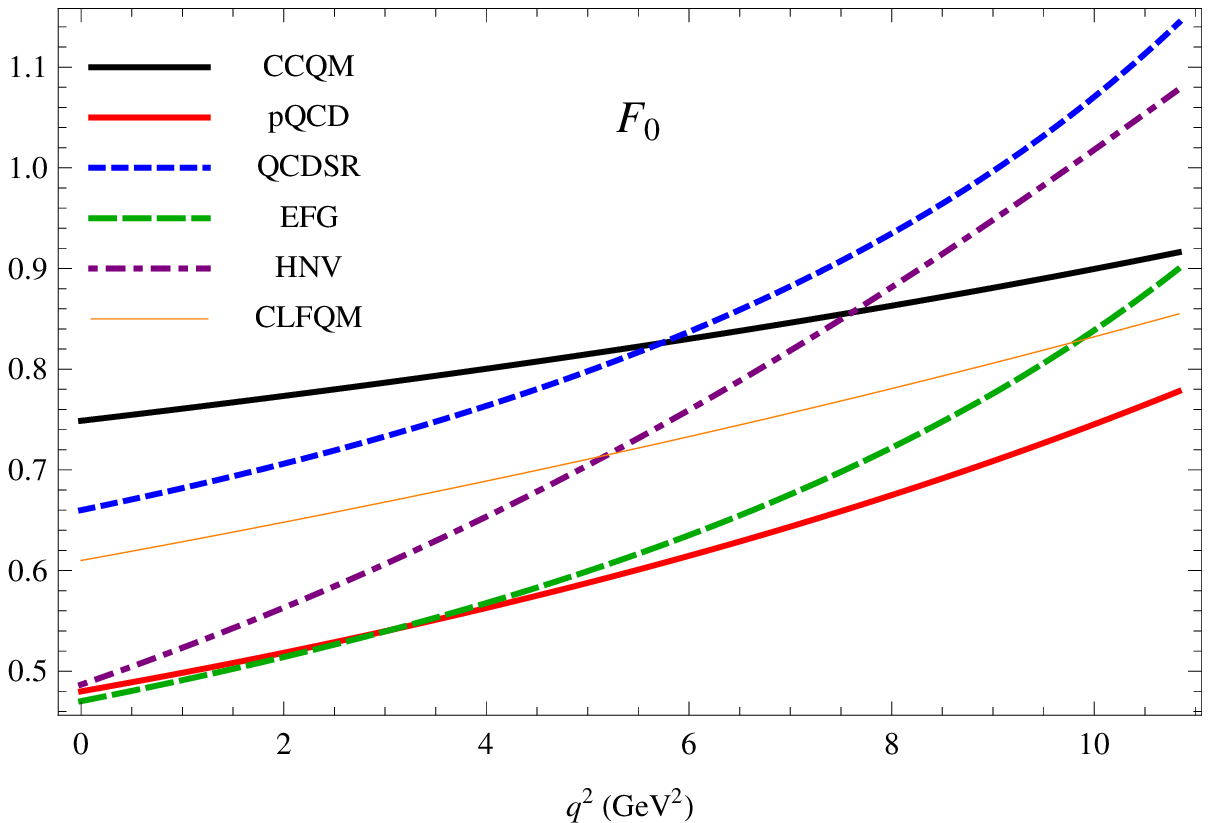}
\end{tabular}
\caption{Comparison of our form factors (CCQM) for the $B_c\to\eta_c$ transition with those from Ref.~\cite{Wen-Fei:2013uea} (pQCD), Ref.~\cite{Kiselev:2002vz} (QCDSR), Ref.~\cite{Ebert:2003cn} (EFG), Ref.~\cite{Hernandez:2006gt} (HNV), and Ref.~\cite{Wang:2008xt} (CLFQM).}
\label{fig:FFEtac-Comp}
\end{figure}

\begin{figure}[htbp]
\begin{tabular}{lr}
\includegraphics[scale=0.6]{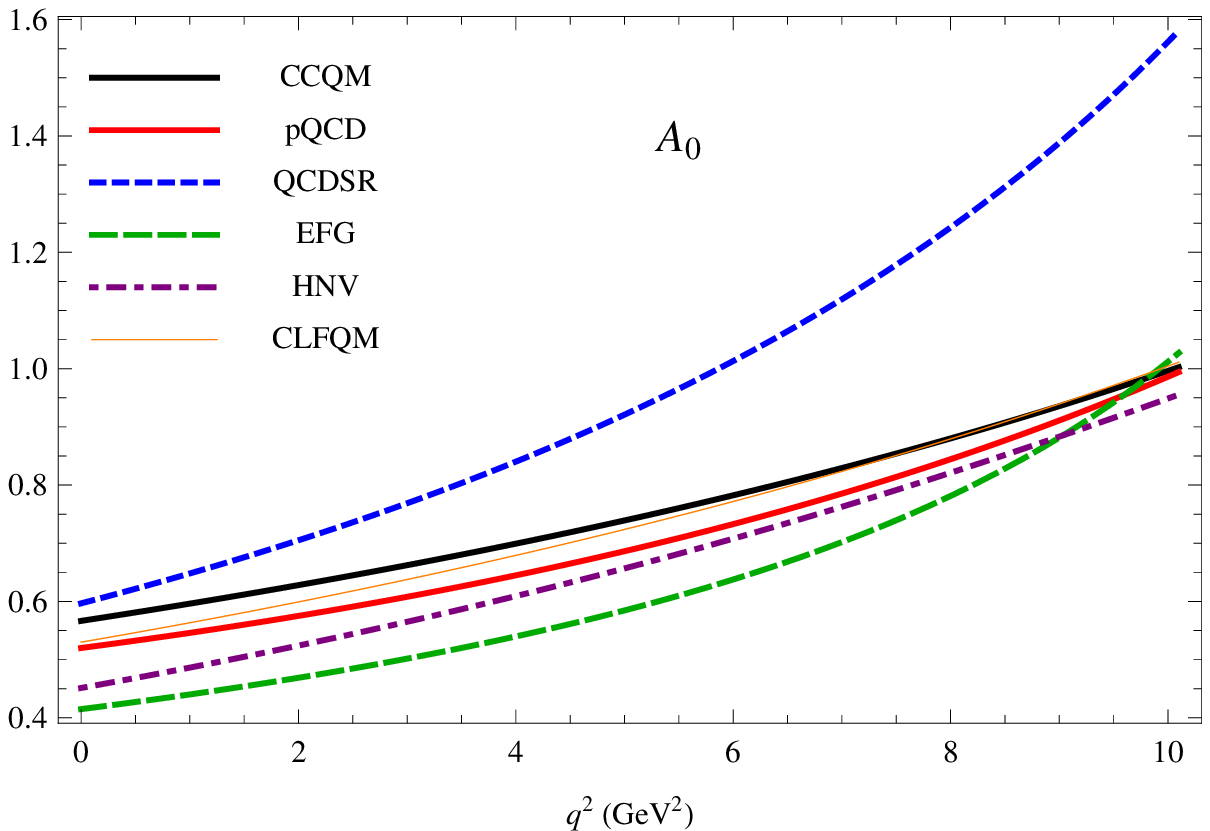}
& 
\includegraphics[scale=0.6]{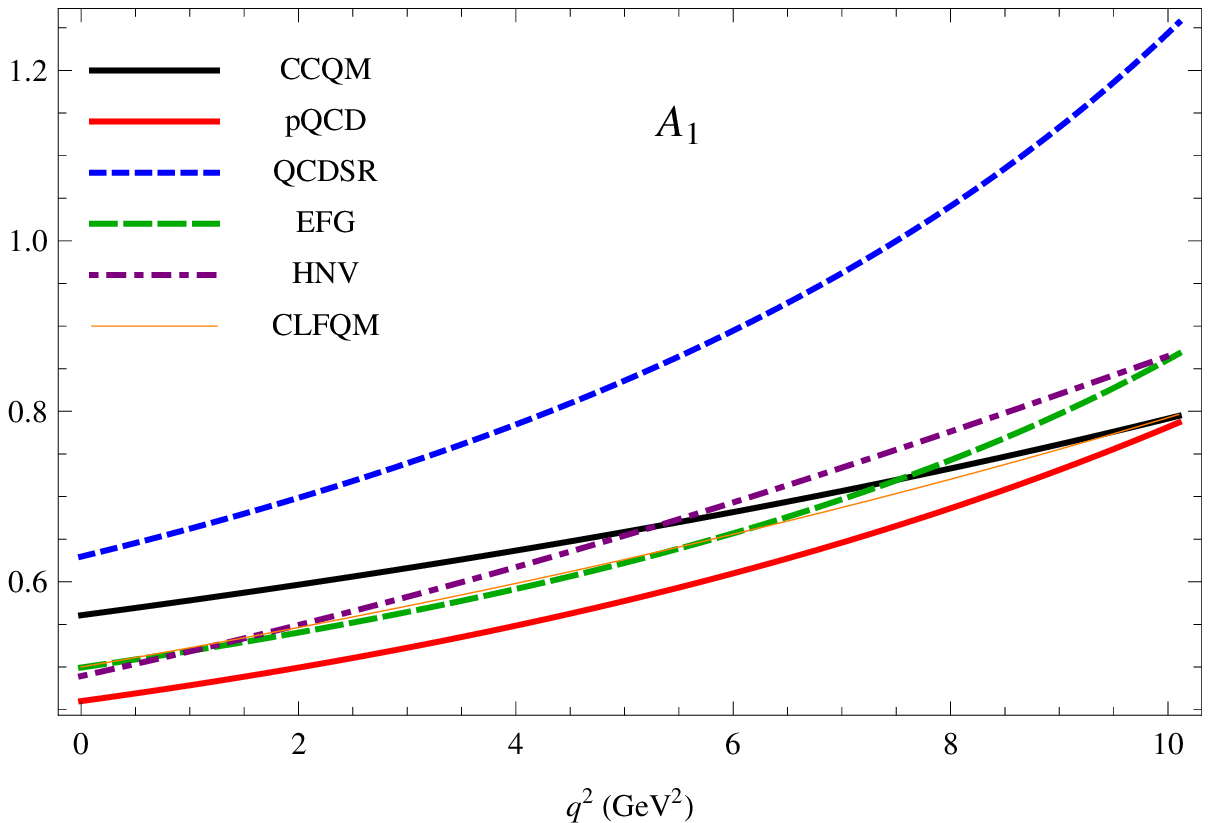}\\
\includegraphics[scale=0.6]{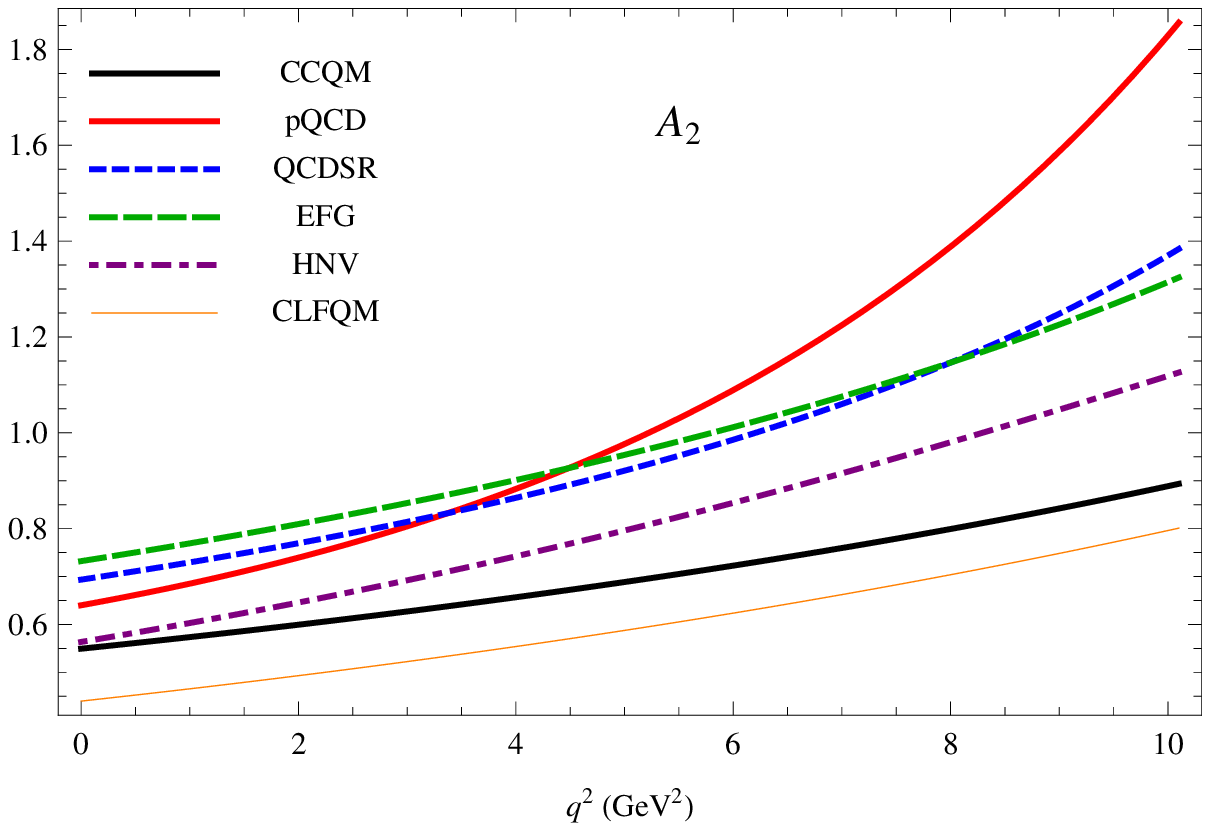}
& 
\includegraphics[scale=0.6]{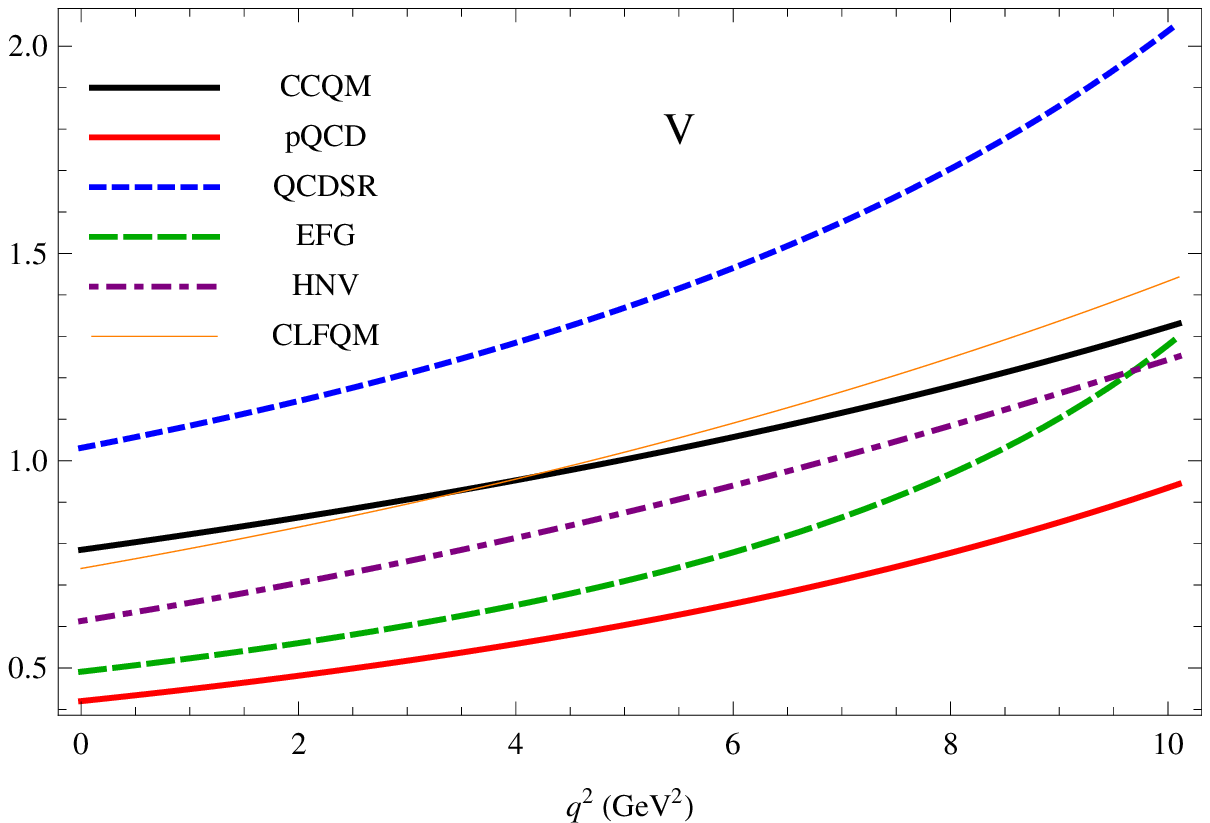}
\end{tabular}
\caption{Comparison of our form factors (CCQM) for the $B_c\to J/\psi$ transition with those from Ref.~\cite{Wen-Fei:2013uea} (pQCD), Ref.~\cite{Kiselev:2002vz} (QCDSR), Ref.~\cite{Ebert:2003cn} (EFG), Ref.~\cite{Hernandez:2006gt} (HNV), and Ref.~\cite{Wang:2008xt} (CLFQM).}
\label{fig:FFJpsi-Comp}
\end{figure}

The $B_c \to J/\psi(\eta_c)$ form factors have been widely calculated in the literature. For a better overview of existed results we perform a comparison between various approaches. For easy comparison, we relate all form factors to the well-known Bauer-Stech-Wirbel form factors~\cite{Wirbel:1985ji}, namely, $F_{+,0}$ for $B_c \to \eta_c$, and $A_{0,1,2}$ and $V$ for $B_c \to J/\psi$.  Note that in Ref.~\cite{Wirbel:1985ji} the notation $F_1$ was used instead of $F_+$. In Fig.~\ref{fig:FFEtac-Comp} and Fig.~\ref{fig:FFJpsi-Comp} we compare our form factors with those obtained in other approaches, namely, perturbative QCD~\cite{Wen-Fei:2013uea}, QCD sum rules (QCDSR)~\cite{Kiselev:2002vz}, the Ebert-Faustov-Galkin relativistic quark model~\cite{Ebert:2003cn}, the Hernandez-Nieves-Verde-Velasco (HNV) nonrelativistic quark model~\cite{Hernandez:2006gt}, and the covariant light-front quark model (CLFQM)~\cite{Wang:2008xt}. It is interesting to note that our form factors are very close to those computed in the CLFQM~\cite{Wang:2008xt}.

\begin{figure}[htbp]
\includegraphics[scale=0.6]{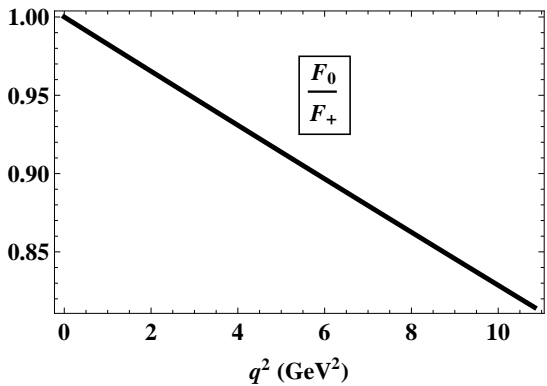}
\caption{Linear behavior of the ratio $F_0(q^2)/F_+(q^2)$.}
\label{fig:F0Fp-ratio}
\end{figure}

Using the heavy quark spin symmetry, the authors of Ref.~\cite{Kiselev:1999sc} have obtained several relations between the form factors of the $B_c\to J/\psi(\eta_c)$ transitions. In particular, the relation between the form factors $F_+$ and $F_-$ can be used to prove the linear behavior of the ratio $F_0(q^2)/F_+(q^2)$,
\be
F_0(q^2)=F_+(q^2)+\frac{q^2}{Pq}F_-(q^2),\qquad
\frac{F_0(q^2)}{F_+(q^2)}=1-\alpha q^2,
\en
where the slope $\alpha$ only depends on the masses of the involved quarks and hadrons. We find $\alpha=0.020$ 
GeV$^{-2}$ from the numerical values in Ref.~\cite{Kiselev:1999sc}. Similarly, we obtain $\alpha=0.018$ 
GeV$^{-2}$ from results of Refs.~\cite{Wen-Fei:2013uea, Kiselev:2002vz}, $\alpha=0.021$ 
GeV$^{-2}$ from Ref.~\cite{Wang:2008xt}. However,  Ref.~\cite{Ebert:2003cn} and Ref.~\cite{Hernandez:2006gt} yield much smaller values, which are $\alpha=0.005$ GeV$^{-2}$ and $\alpha=0.007$ GeV$^{-2}$, respectively. In our model, the ratio $F_0(q^2)/F_+(q^2)$ exhibits an almost linear behavior in the whole $q^2$ range as demonstrated in Fig.~\ref{fig:F0Fp-ratio}, from which we obtain $\alpha=0.017$ GeV$^{-2}$. The value of the slope $\alpha$ plays an important role in studying the shape of the form factors, which can be determined more accurately by future lattice calculations.

It is also worth mentioning the very recent lattice results for the $B_c \to J/\psi$ form factors provided by the HPQCD Collaboration~\cite{Colquhoun:2016osw}. In this study, they found $A_1(0)=0.49$, $A_1(q^2_{\rm max})=0.79$, and $V(0)=0.77$, which are very close to our values $A_1(0)=0.56$, $A_1(q^2_{\rm max})=0.79$, and $V(0)=0.78$.

Almost all the recent studies on possible NP in the decays $B_c \to (J/\psi,\eta_c)\tau\nu$ employ the form factors $F_{0,+}$, $A_{0,1,2}$, and $V$ calculated in pQCD approach~\cite{Wen-Fei:2013uea}. The remaining form factors corresponding to the NP operators are obtained by using the quark-level equations of motion (EOMs). In this paper we provide the full set of form factors in the SM as well as in the presence of NP operators without relying on the EOMs. However, this does not mean that our form factors do not satisfy the EOMs. A brief discussion of the EOMs in our model can be found in Ref.~\cite{Ivanov:2016qtw}. Our form factors therefore can be used to analyze NP effects in the decays $B_c \to (J/\psi,\eta_c)\tau\nu$ in a self-consistent manner, and independently from other studies.
\section{Experimental constraints}
\label{sec:constraint}
Constraints on the Wilson coefficients appearing in the effective Hamiltonian Eq.~(\ref{eq:Heff}) are obtained by using experimental data for the ratios of branching fractions $R_D=0.407\pm 0.046$, $R_{D^\ast}=0.304\pm 0.015$~\cite{Amhis:2016xyh}, and $R_{J/\psi}=0.71\pm 0.25$~\cite{Aaij:2017tyk}, as well as the requirement $\mathcal{B}(B_c\to \tau\nu)\leq 10\%$ from the LEP1 data~\cite{Akeroyd:2017mhr}. It should be mentioned that within the SM our calculation yields $R_D=0.267$, $R_{D^\ast}=0.238$, and $R_{J/\psi}=0.24$. We take into account a theoretical error of $10\%$ for our ratios. Besides, we assume the dominance of only one NP operator besides the SM contribution, which means that only one NP Wilson coefficient is considered at a time.
\begin{figure}[htbp]
\begin{tabular}{cc}
\includegraphics[scale=0.4]{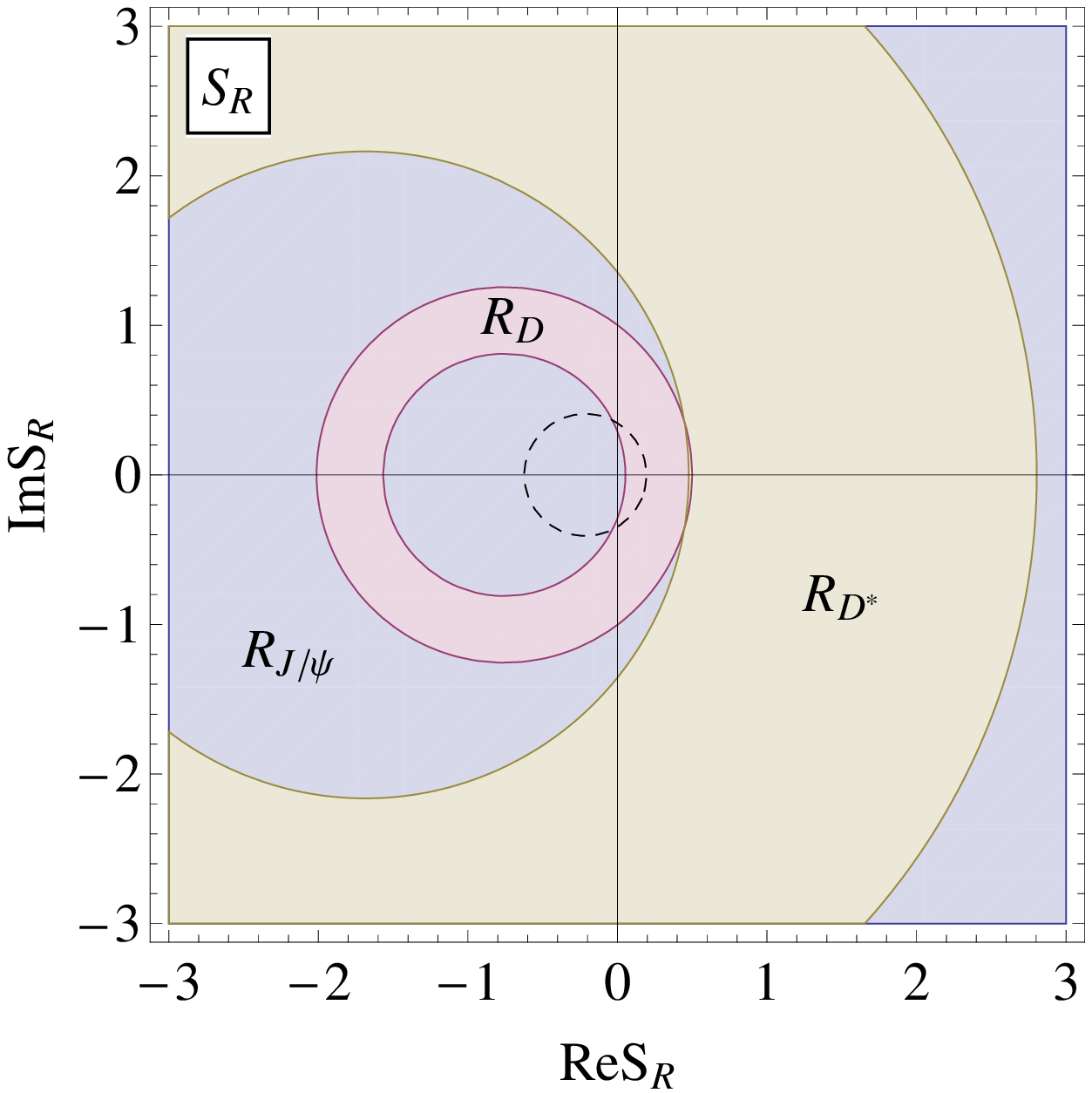}&
\includegraphics[scale=0.4]{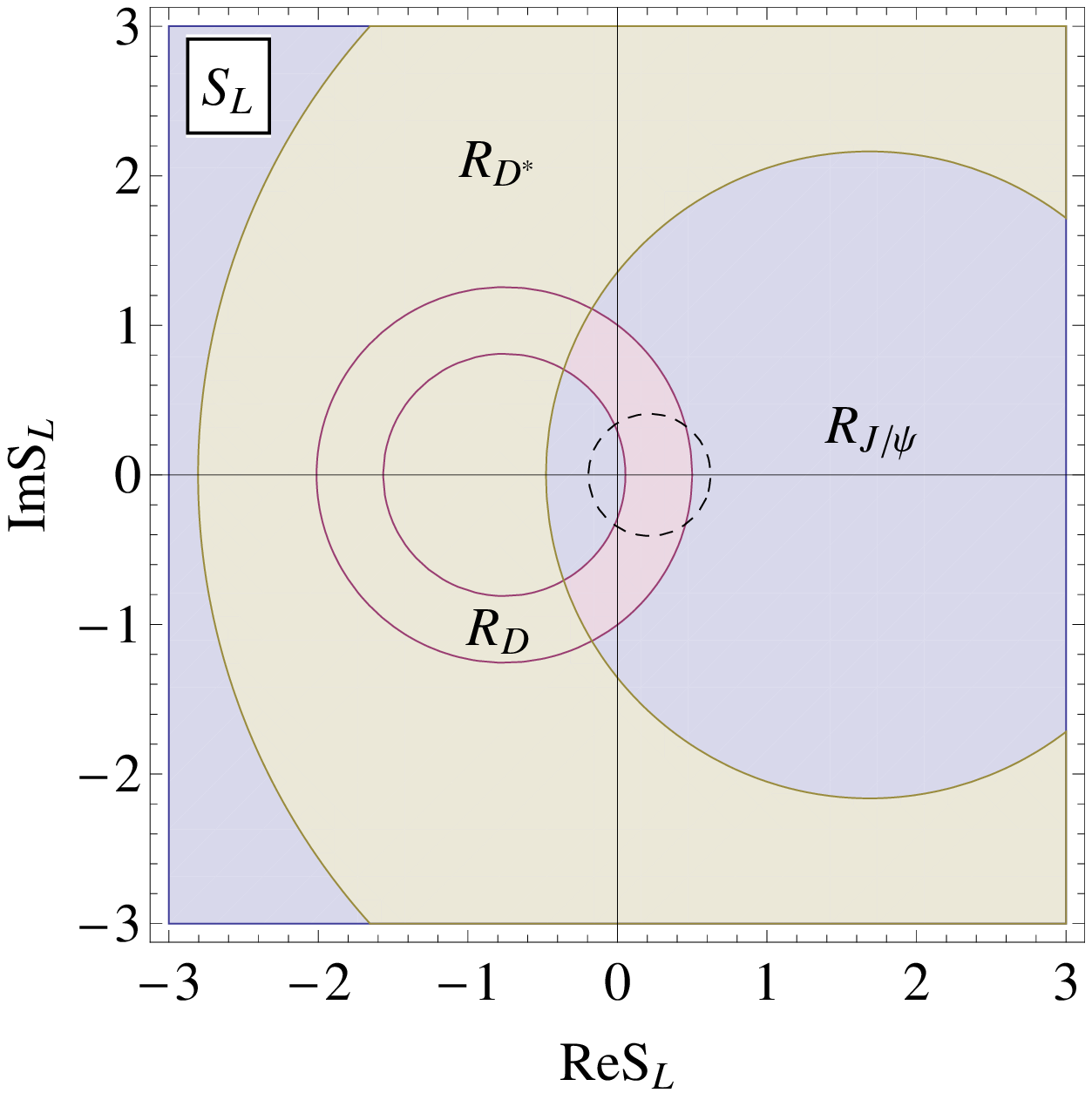}
\end{tabular}
\caption{Constraints on the Wilson coefficients $S_R$ and $S_L$ from the measurements of $R_{J/\psi}$, $R_D$, and $R_{D^\ast}$ within $2\sigma$, and from the branching fraction $\mathcal{B}(B_c \to \tau\nu)$ (dashed curve).}
\label{fig:constraintS}
\end{figure}

In Fig.~\ref{fig:constraintS} we show the constraints on the scalar Wilson coefficients $S_{L,R}$ within $2\sigma$. It is seen that the recent experimental value of $R_{J/\psi}$ does not give any additional constraint on $S_{L,R}$ to what have been obtained by using $R_{D^{(\ast)}}$. In particular, $S_{R}$ is excluded within $2\sigma$ using only $R_{D^{(\ast)}}$. However, in the case of $S_L$, the constraint from $\mathcal{B}(B_c \to \tau\nu)$ plays the main role in ruling out $S_L$. In general, the branching of the tauonic $B_c$ decay imposes a severe constraint on the scalar NP scenarios. Many models of NP involving new particles, such as charged Higgses or leptoquarks, also suffer from the same constraint, and therefore need additional modifications to accommodate the current experimental data (see e.g. Refs.~\cite{Crivellin:2017zlb, Lee:2017kbi, Iguro:2017ysu}).
\begin{figure}[htbp]
\begin{tabular}{ccc}
\includegraphics[scale=0.4]{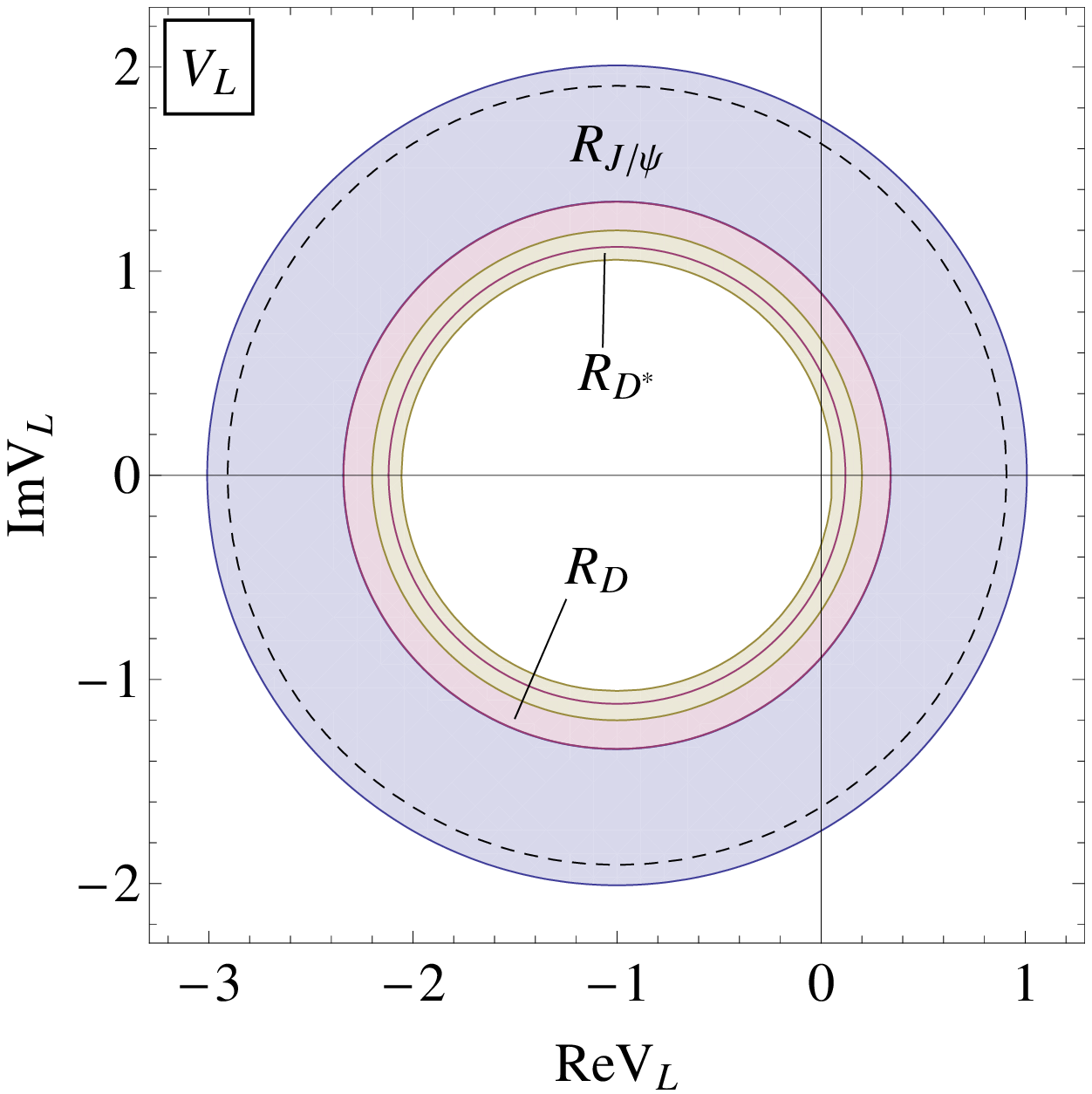}&
\includegraphics[scale=0.4]{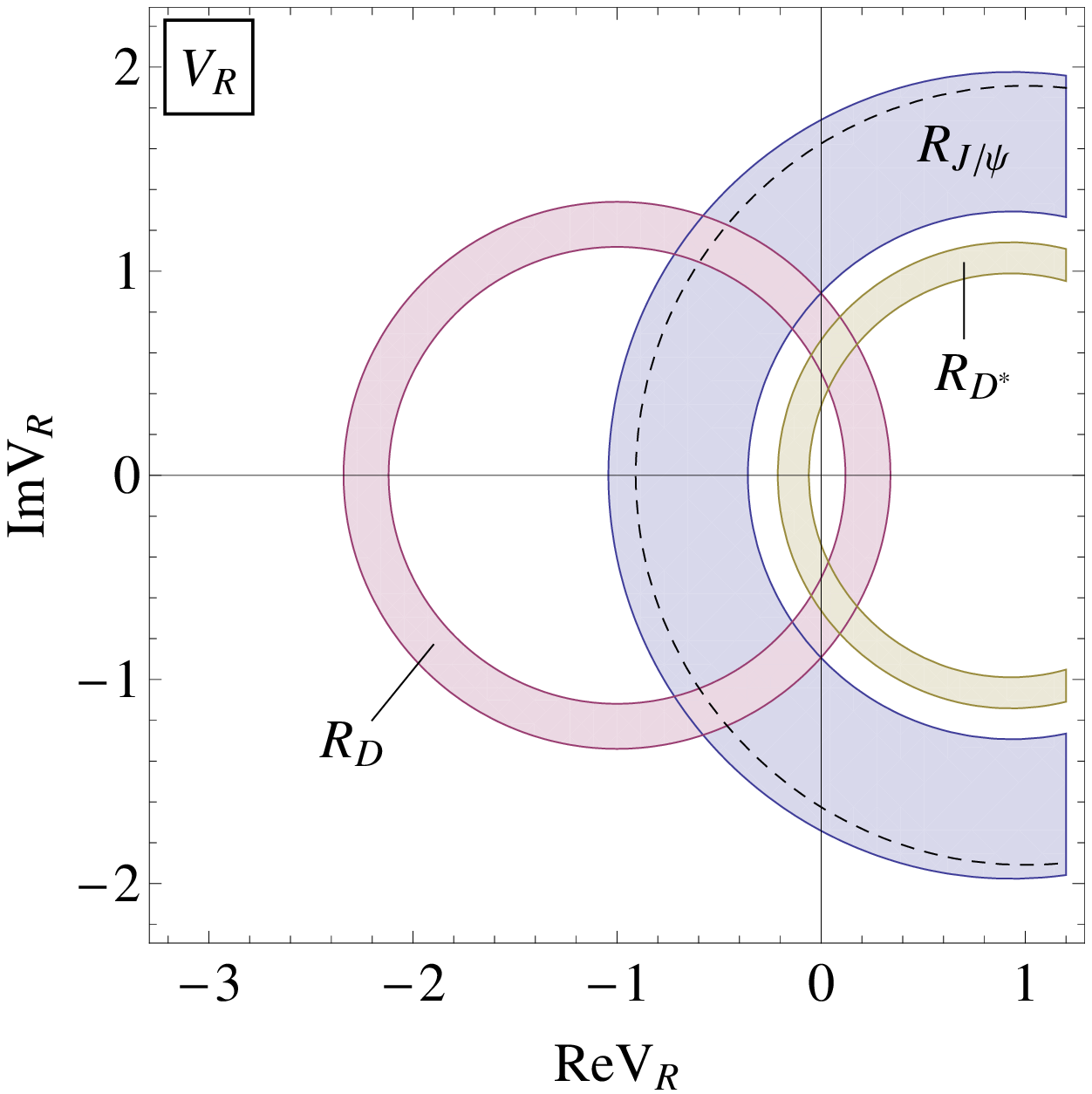}
& 
\includegraphics[scale=0.4]{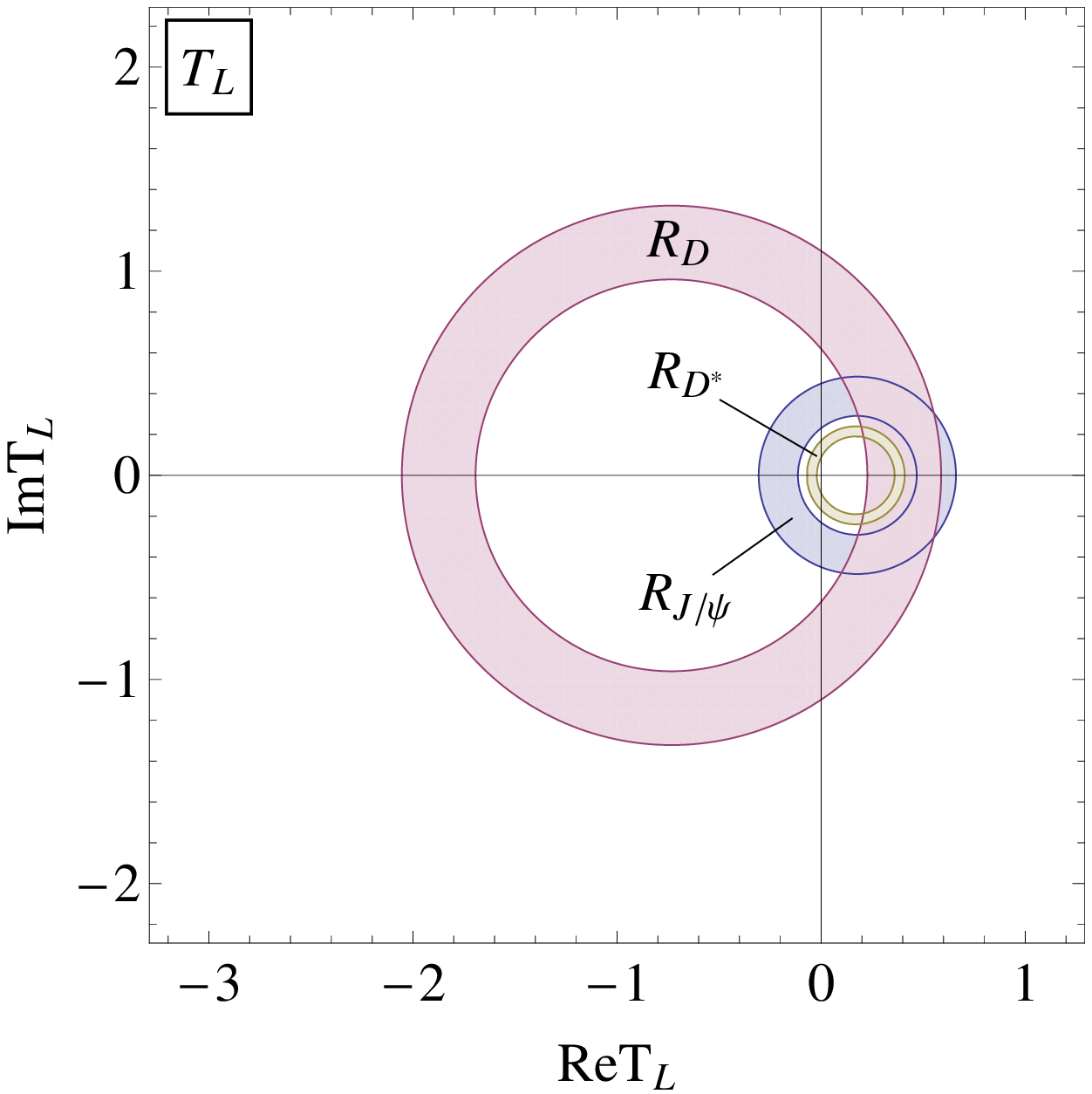}
\\
\includegraphics[scale=0.4]{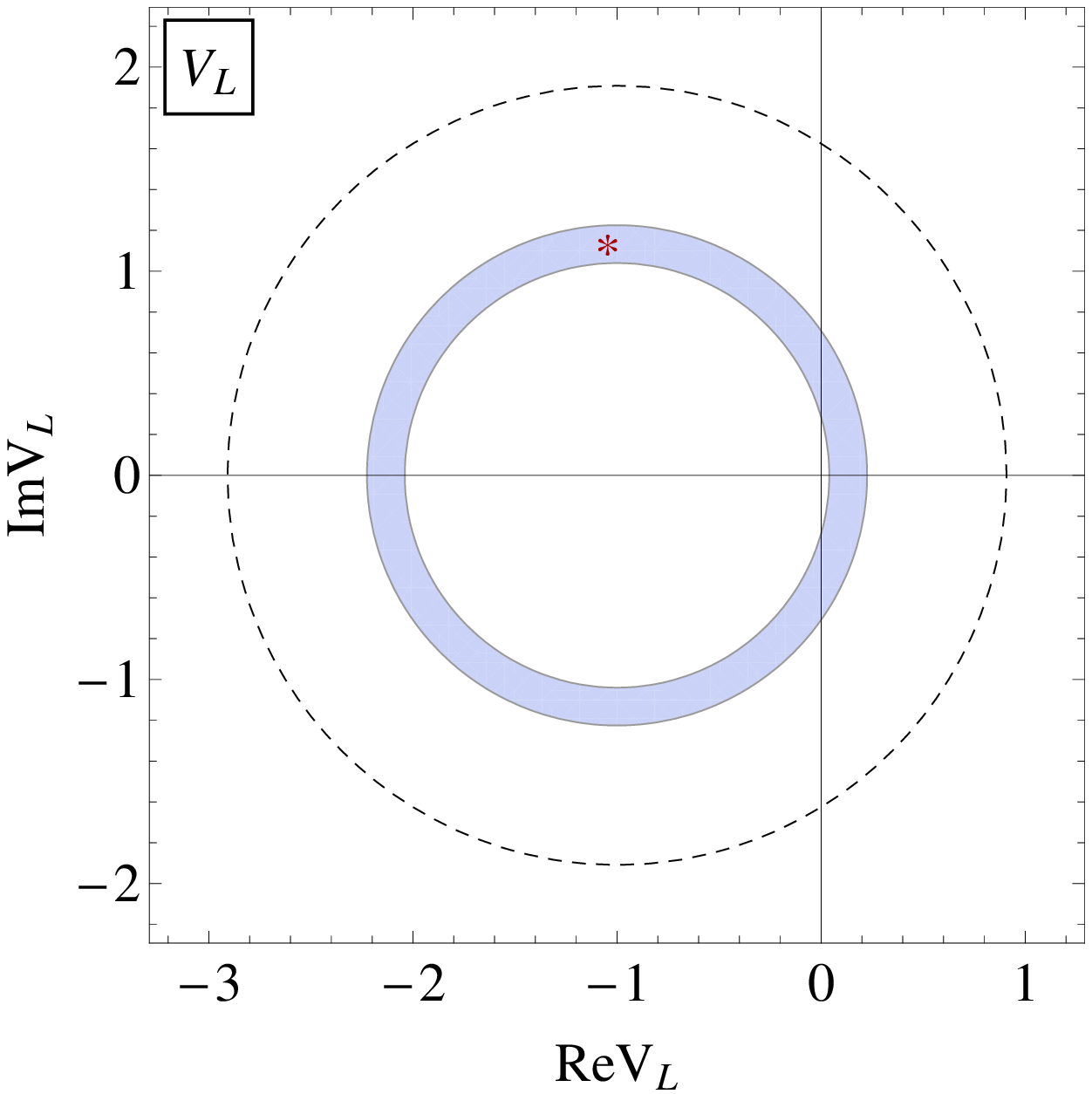}
&
\includegraphics[scale=0.4]{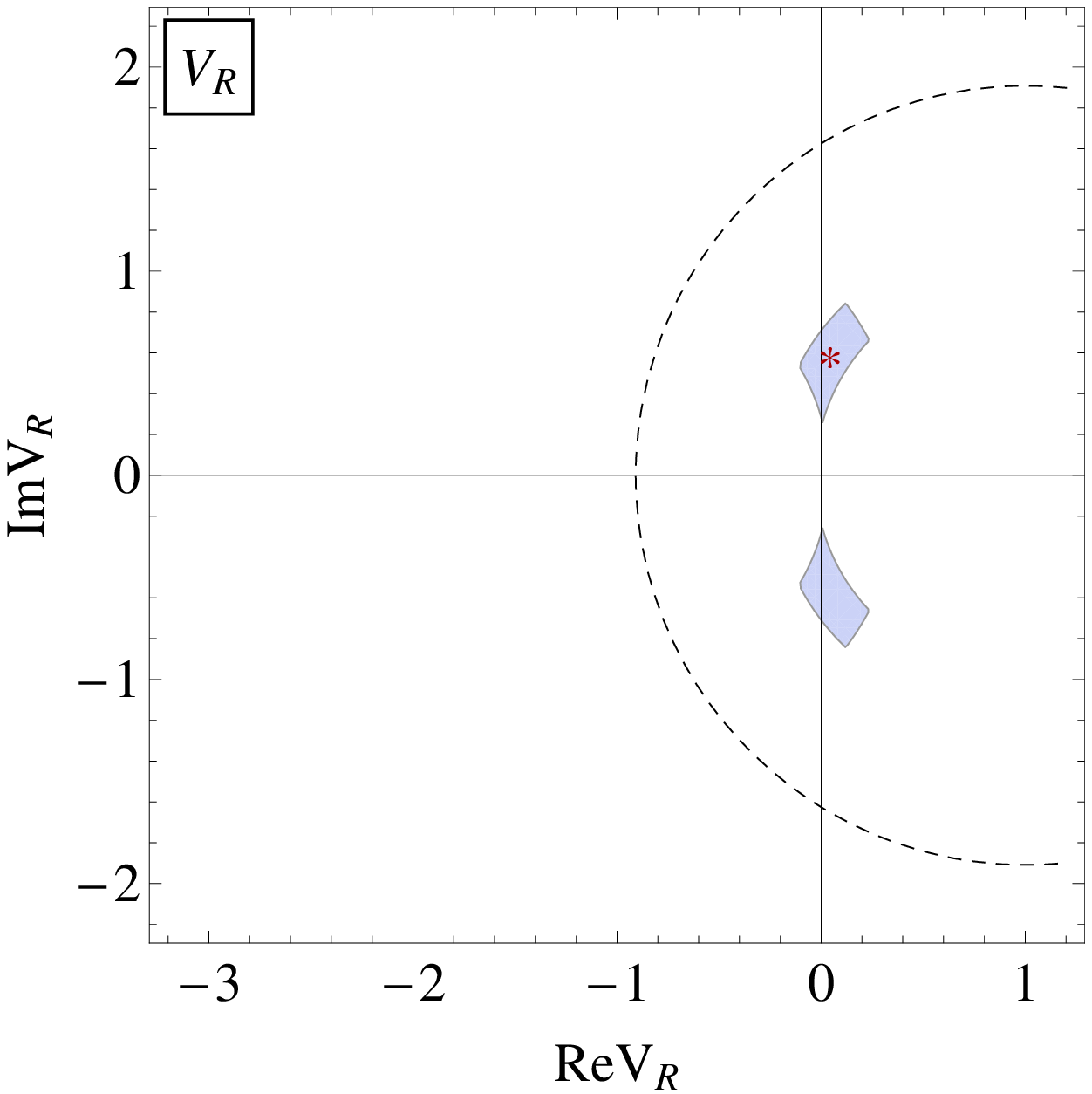}
&
\includegraphics[scale=0.4]{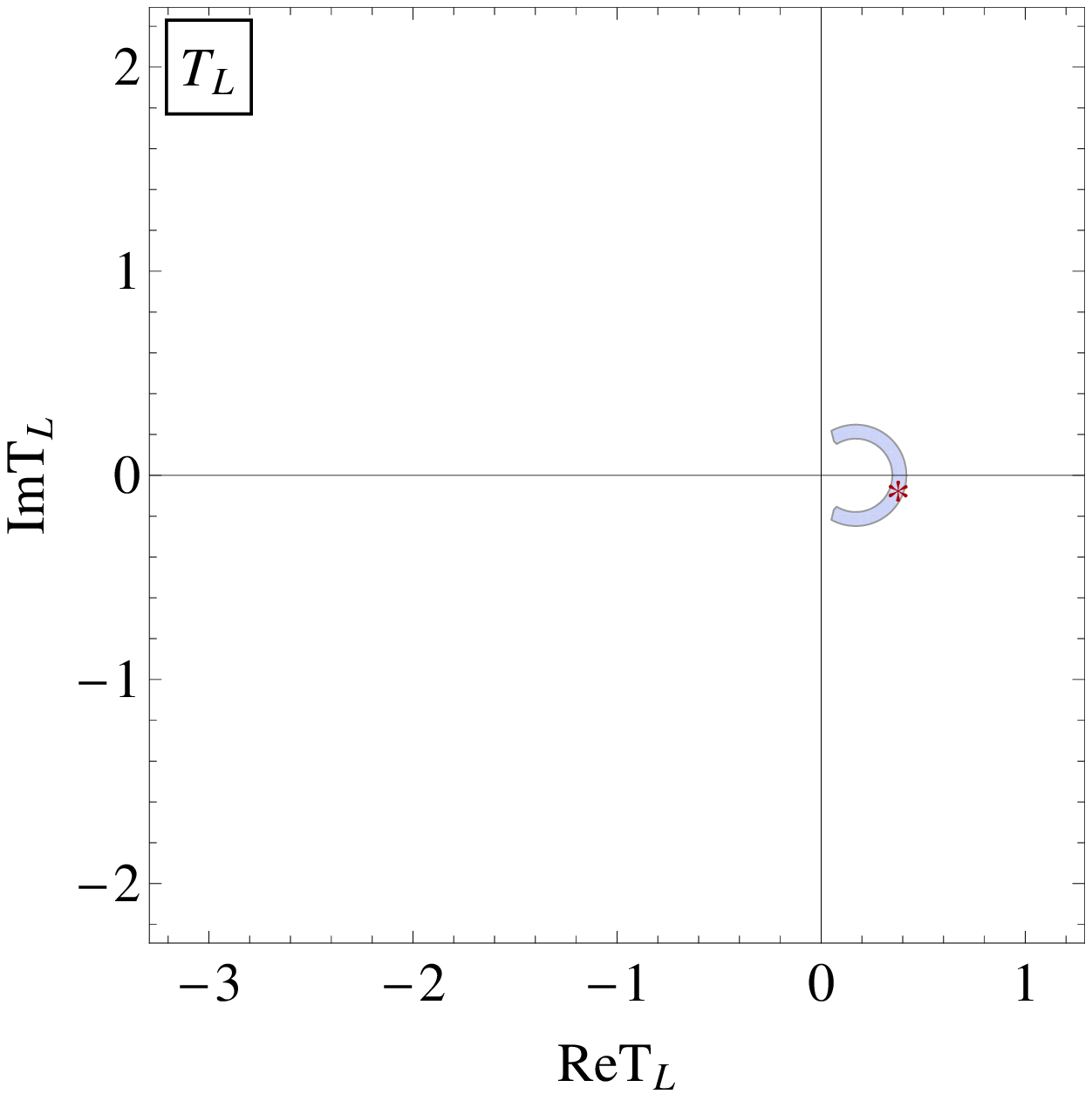}
\end{tabular}
\caption{Constraints on the Wilson coefficients $V_R$, $V_L$, and $T_L$ from the measurements of $R_{J/\psi}$, $R_D$, and $R_{D^\ast}$ within $1\sigma$ (upper panels) and $2\sigma$ (lower panels), and from the branching fraction $\mathcal{B}(B_c \to \tau\nu)$ (dashed curve).}
\label{fig:constraintVT}
\end{figure}

In the upper panels of Fig.~\ref{fig:constraintVT} we present the constraints on the vector $V_{L,R}$ and tensor $T_L$ Wilson coefficients. There is no available space for these coefficients within $1\sigma$. Moreover, they are excluded mainly due to the additional constraint from $R_{J/\psi}$, rather than from $\mathcal{B}(B_c \to \tau\nu)$. This holds exactly in the case of $T_L$ since the operator $\mathcal{O}_{T_L}$ has no effect on $\mathcal{B}(B_c \to \tau\nu)$. In the lower panels of 
Fig.~\ref{fig:constraintVT} we show the allowed regions for $V_{L,R}$ and $T_L$ within $2\sigma$. In each allowed region at $2\sigma$ we find a best-fit value for each NP coupling. The best-fit couplings read $V_L =-1.05+i1.15$, $V_R =0.04+i0.60$, $T_L =0.38-i0.06$, and are marked with an asterisk.
\section{Theoretical predictions}
\label{sec:prediction}
\begin{figure}[htbp]
\begin{center}
\begin{tabular}{ccc}
\includegraphics[scale=0.4]{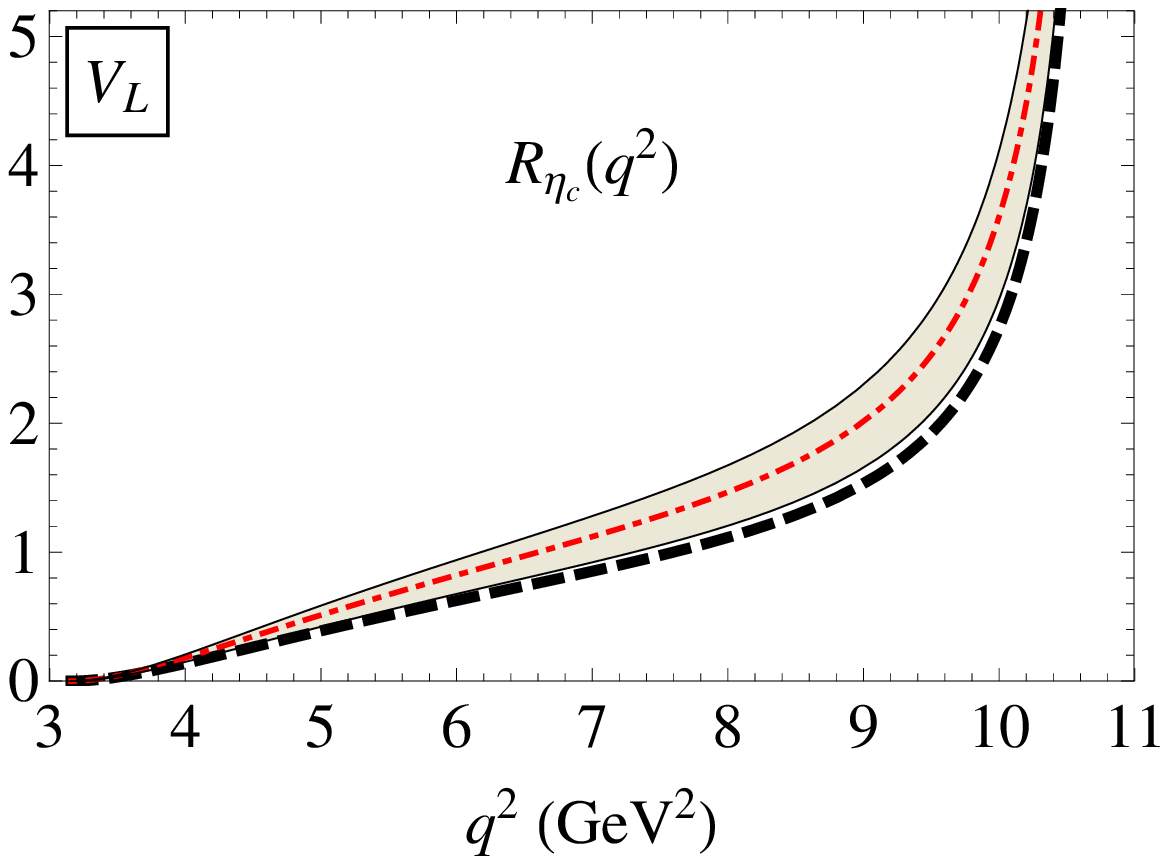}&
\includegraphics[scale=0.4]{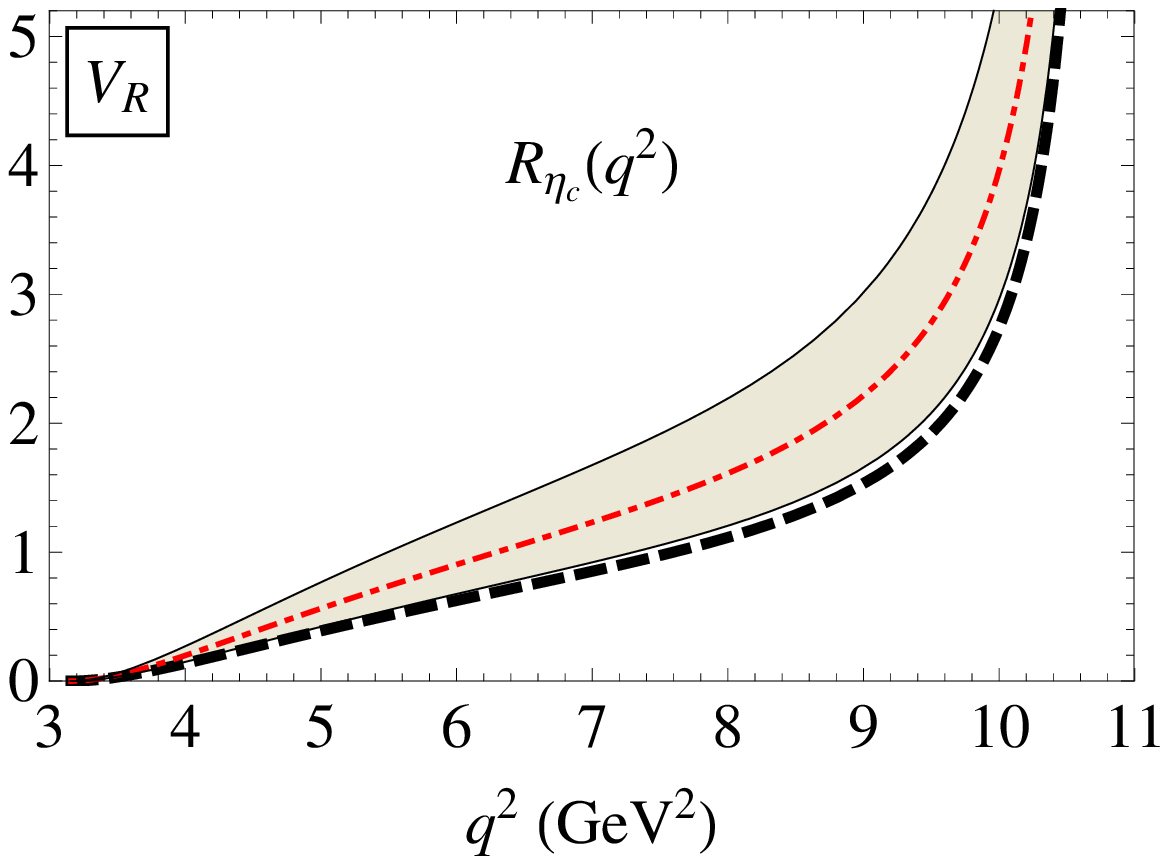}&
\includegraphics[scale=0.4]{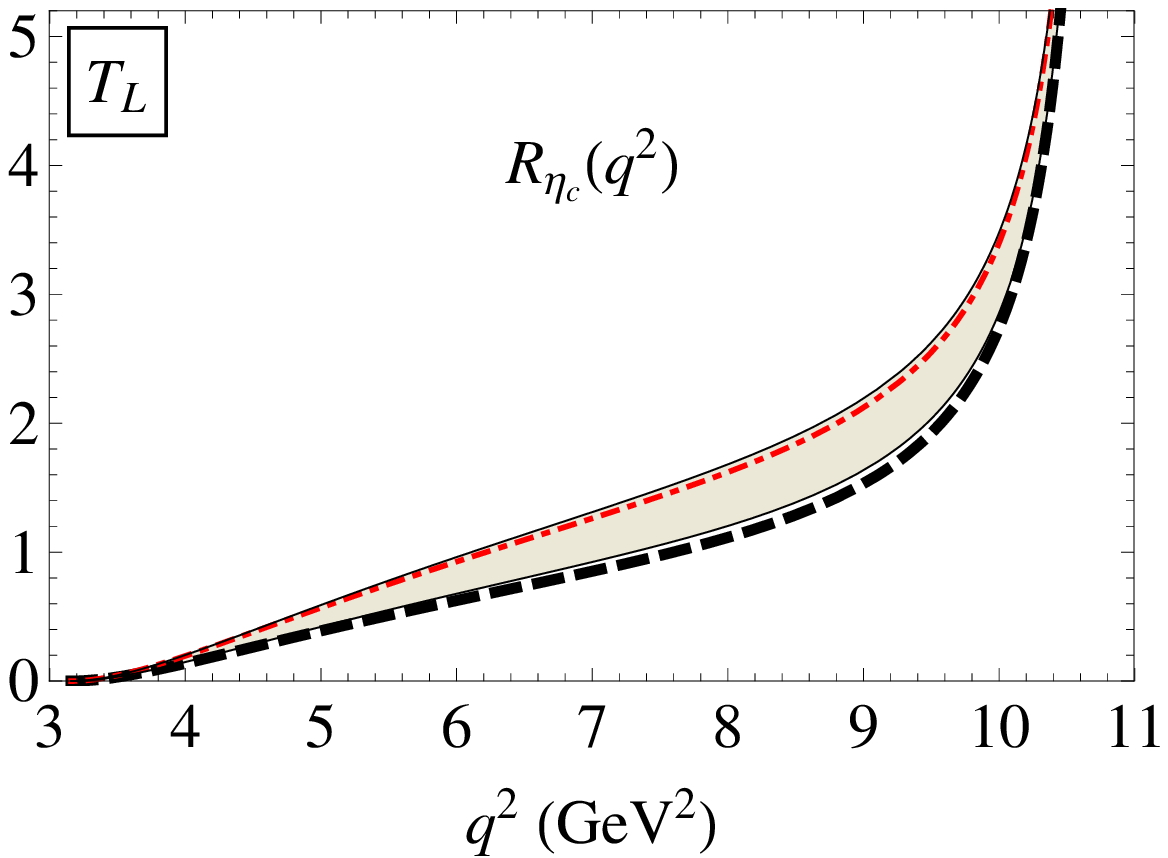}\\
\includegraphics[scale=0.4]{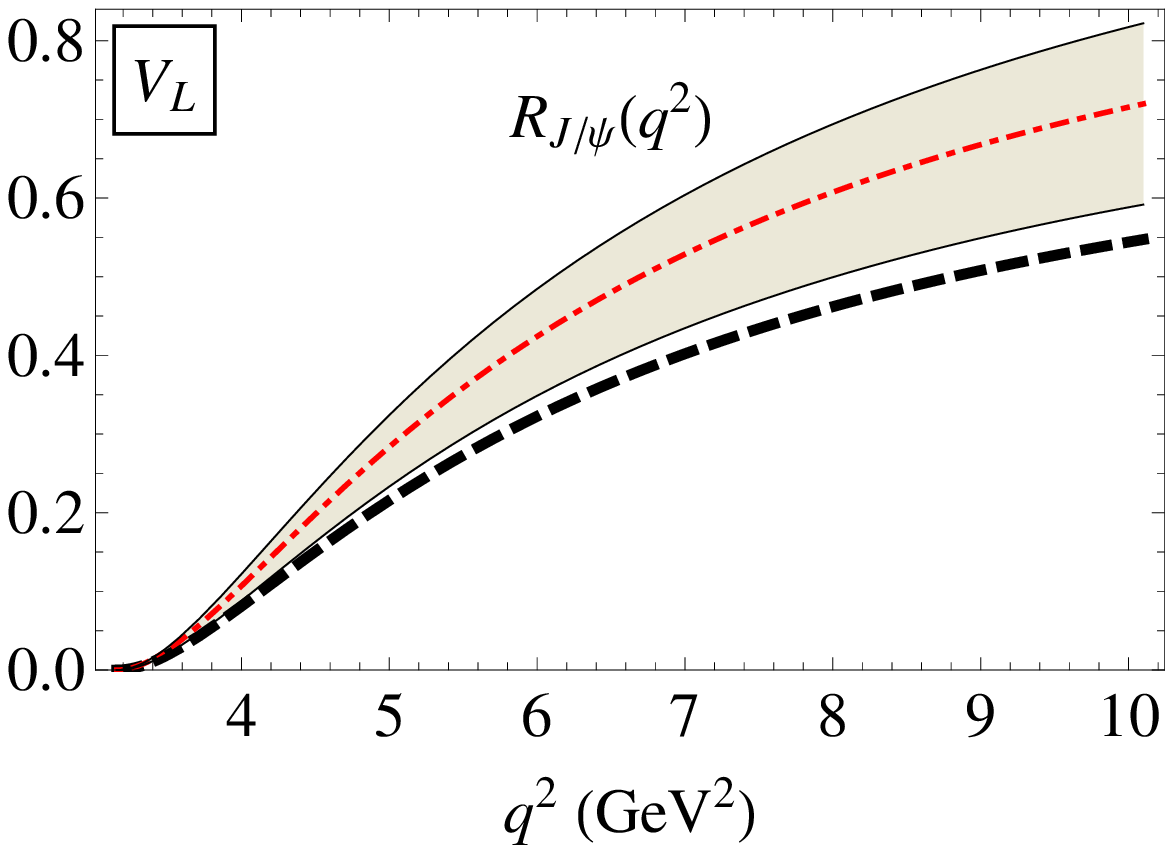}&
\includegraphics[scale=0.4]{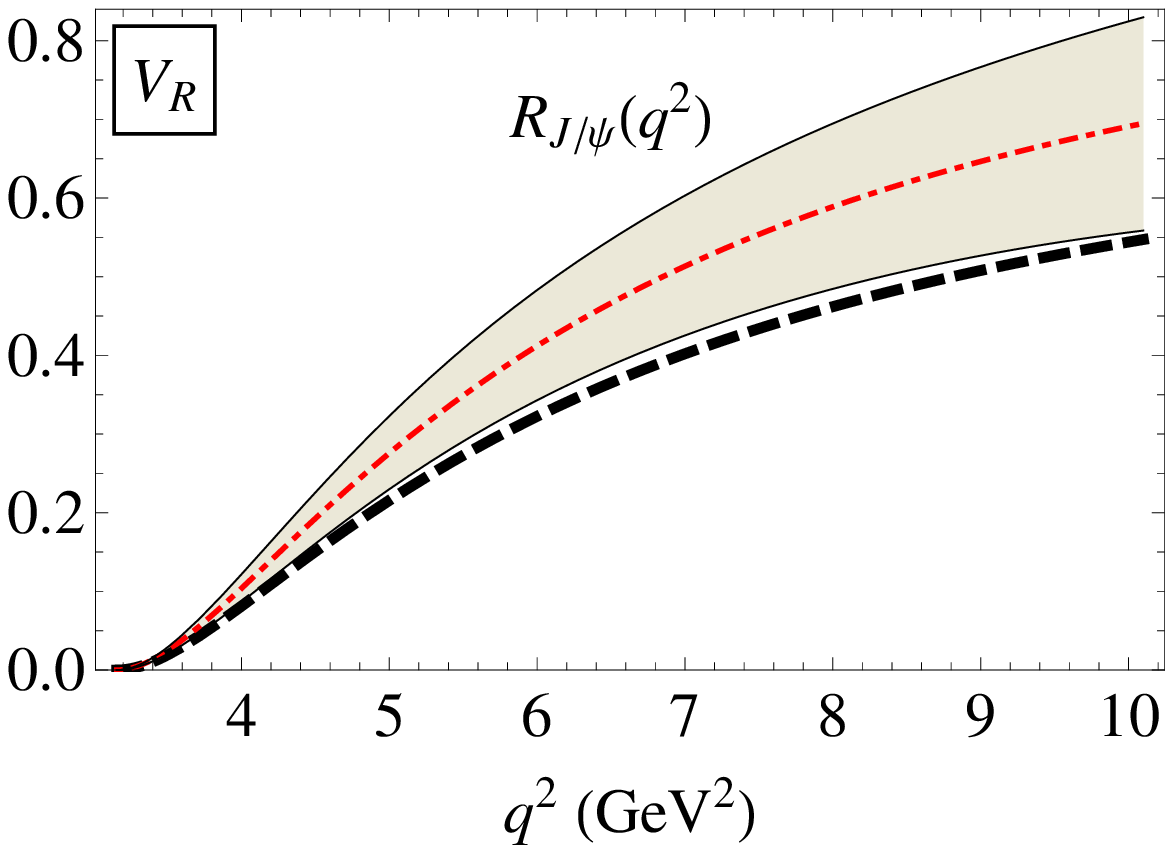}&
\includegraphics[scale=0.4]{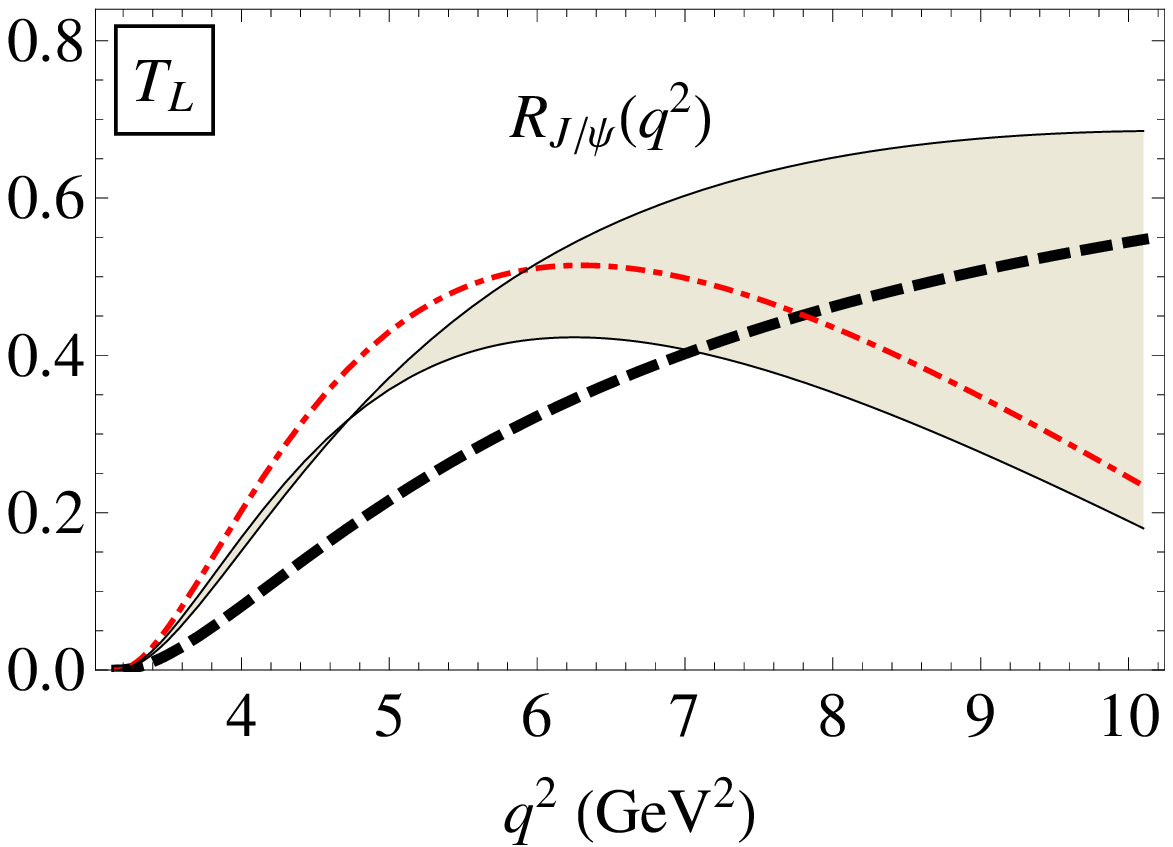}
\end{tabular}
\caption{Differential ratios $R_{\eta_c}(q^2)$ (upper panels) and $R_{J/\psi}(q^2)$ (lower panels). The thick black dashed lines are the SM prediction; the gray bands include NP effects corresponding to the $2\sigma$ allowed regions in Fig.~\ref{fig:constraintVT}; the red dot-dashed lines represent the best-fit values of the NP couplings.}
\label{fig:R}
\end{center}
\end{figure}

In this section we use the $2\sigma$ allowed regions for $V_{L,R}$ and $T_L$ obtained in the previous section to analyze their effects on several physical observables. Firstly, in Fig.~\ref{fig:R} we show the $q^2$ dependence of the ratios $R_{J/\psi}$ and $R_{\eta_c}$ in different NP scenarios. It is obvious that all the NP operators $\mathcal{O}_{V_L}$, $\mathcal{O}_{V_R}$, and $\mathcal{O}_{T_L}$ increase the ratios. However, it is interesting to note that $\mathcal{O}_{T_L}$ can change the shape of $R_{J/\psi}(q^2)$ and may imply a peak in the distribution. This unique behavior can help identify the tensor origin of NP by studying the $q^2$ distribution of the decay $B_c\to J/\psi \tau\nu$. 
\begin{table}[htbp] 
\begin{center}
\begin{tabular}{ccc}
\hline\hline
&\quad  $<R_{\eta_c}>$ \qquad 
&\quad  $<R_{J/\psi}>$ \qquad   
\\
\hline
SM &\quad $0.26$\quad &\quad $0.24$\quad \\
$V_L$
&\quad $(0.28,0.39)$\quad
&\quad $(0.26,0.37)$\quad
\\
$V_R$
&\quad $(0.28,0.51)$\quad
&\quad $(0.26,0.37)$\quad
\\
$T_L$
&\quad $(0.28,0.38)$\quad
&\quad $(0.24,0.36)$\quad
\\
\hline\hline
\end{tabular}
\caption{The $q^2$ average of the ratios in the SM and in the presence of NP.}
\label{tab:ratio}
\end{center}
\end{table}

The average values of the ratios $R_{J/\psi}$ and $R_{\eta_c}$ over the whole $q^2$ region are given in Table~\ref{tab:ratio}. The row labeled by SM contains our predictions within the SM using our form factors. The predicted ranges for the ratios in the presence of NP are given in correspondence with the $2\sigma$ allowed regions of the NP couplings shown in 
Fig.~\ref{fig:constraintVT}. Here, the most visible effect comes from the operator $\mathcal{O}_{V_R}$, which can increase the average ratio $<R_{\eta_c}>$ by a factor of 2.

Next, we consider the polarization observables in these decays. For this purpose we write the differential $(q^2,\cos\theta)$ distribution as
\be
\frac{d\Gamma(B_c\to J/\psi(\eta_c)\tau\nu)}{dq^2d\cos\theta}
=\frac{G_F^2|V_{cb}|^2|{\bf p_2}|q^2}{(2\pi)^3 16m_1^2}\Big(1-\frac{m_\tau}{q^2}\Big)^2\cdot W(\theta),
\en
where $W(\theta)$ is the polar angular distribution, which is described by a
tilted parabola. For convenience we define a normalized polar angular distribution 
$\widetilde W(\theta)$ as follows:
\be
\widetilde W(\theta)=\frac{W(\theta)}
{ {\cal H}_{\rm tot}}=\frac{a+b\cos\theta+c\cos^{2}\theta}{2(a+c/3)}.
\label{eq:normdis}
\en
\noindent 
The normalized angular decay distribution 
$\widetilde W(\theta)$ obviously integrates to $1$ after
$\cos\theta$ integration. 
The linear
coefficient $b/2(a+c/3)$ can be projected out by defining a 
forward-backward asymmetry given by 
\be
\mathcal{A}_{FB}(q^2)=
\frac{d\Gamma(F)-d\Gamma(B)}{d\Gamma(F)+d\Gamma(B)}=
\frac{ [\int_{0}^{1}-\int_{-1}^{0}] d\cos\theta\, d\Gamma/d\cos\theta}
     { [\int_{0}^{1}+\int_{-1}^{0}] d\cos\theta\, d\Gamma/d\cos\theta}
= \frac{b}{2(a+c/3)}.
\label{eq:fbAsym}
\en
The quadratic coefficient $c/2(a+c/3)$ is obtained by taking the second derivative of $\widetilde W(\theta)$. We therefore
define a convexity parameter by writing 
\be
C_F^\tau(q^2) = \frac{d^{2}\widetilde W(\theta)}{d(\cos\theta)^{2}}
= \frac{c}{a+c/3}.
\en

\begin{figure}[htbp]
\begin{center}
\begin{tabular}{ccc}
\includegraphics[scale=0.4]{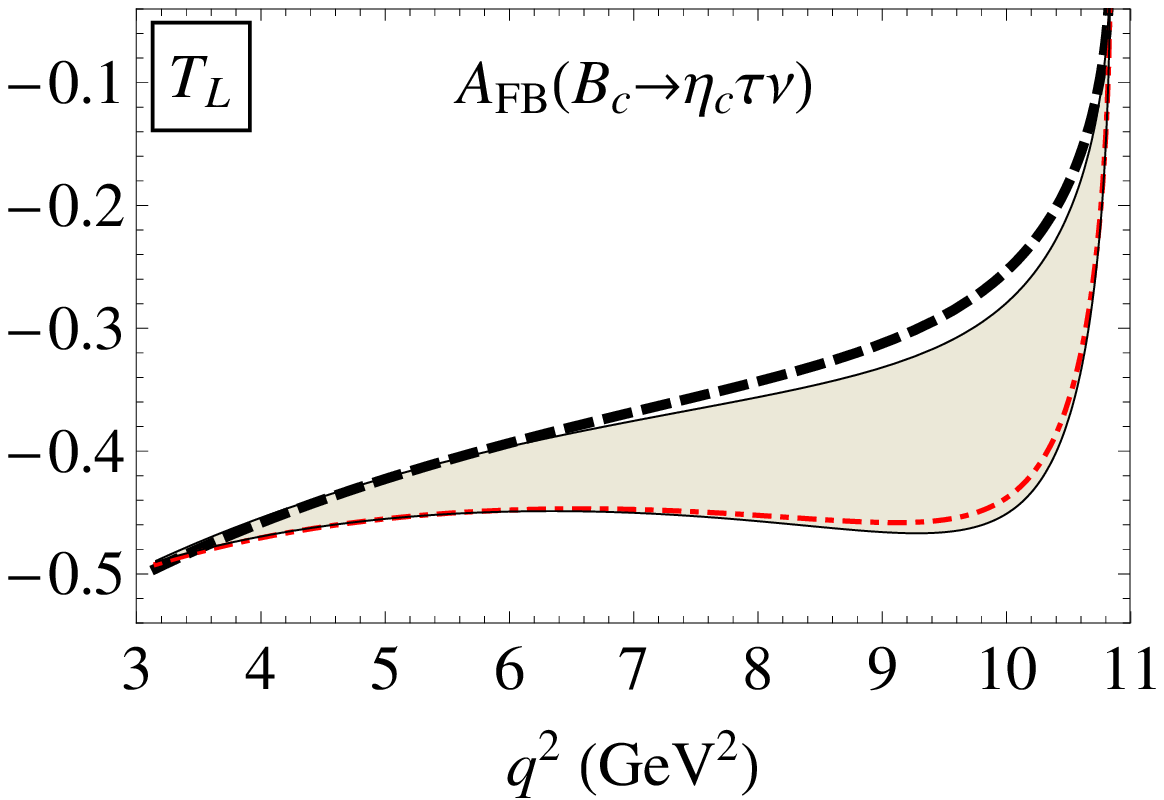}&
\includegraphics[scale=0.4]{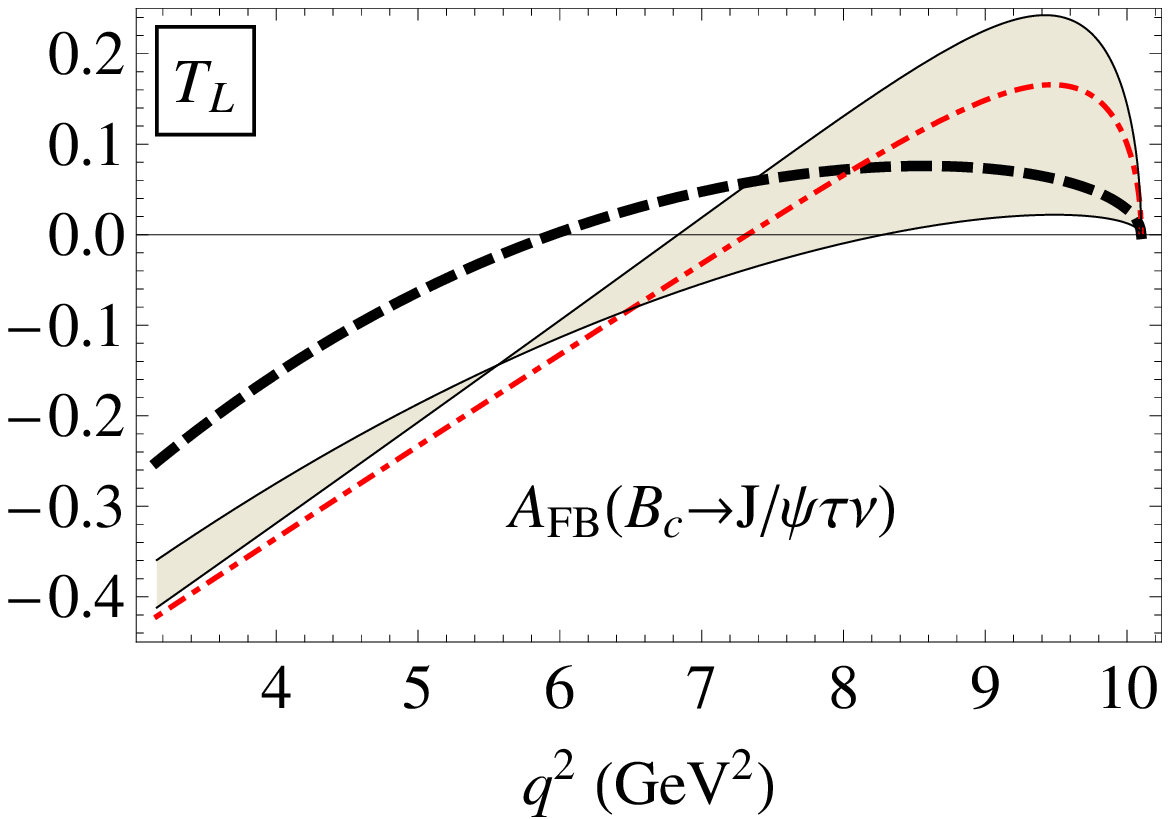}&
\includegraphics[scale=0.4]
{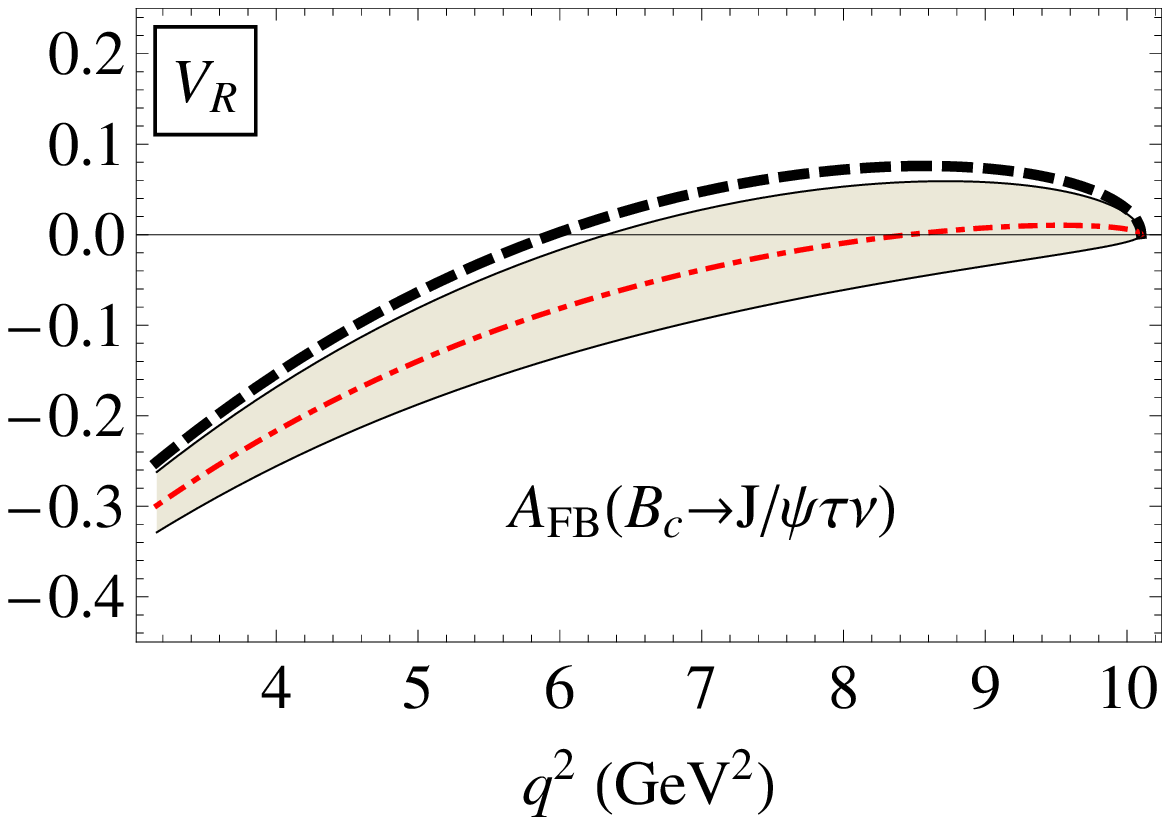}\\
\includegraphics[scale=0.4]{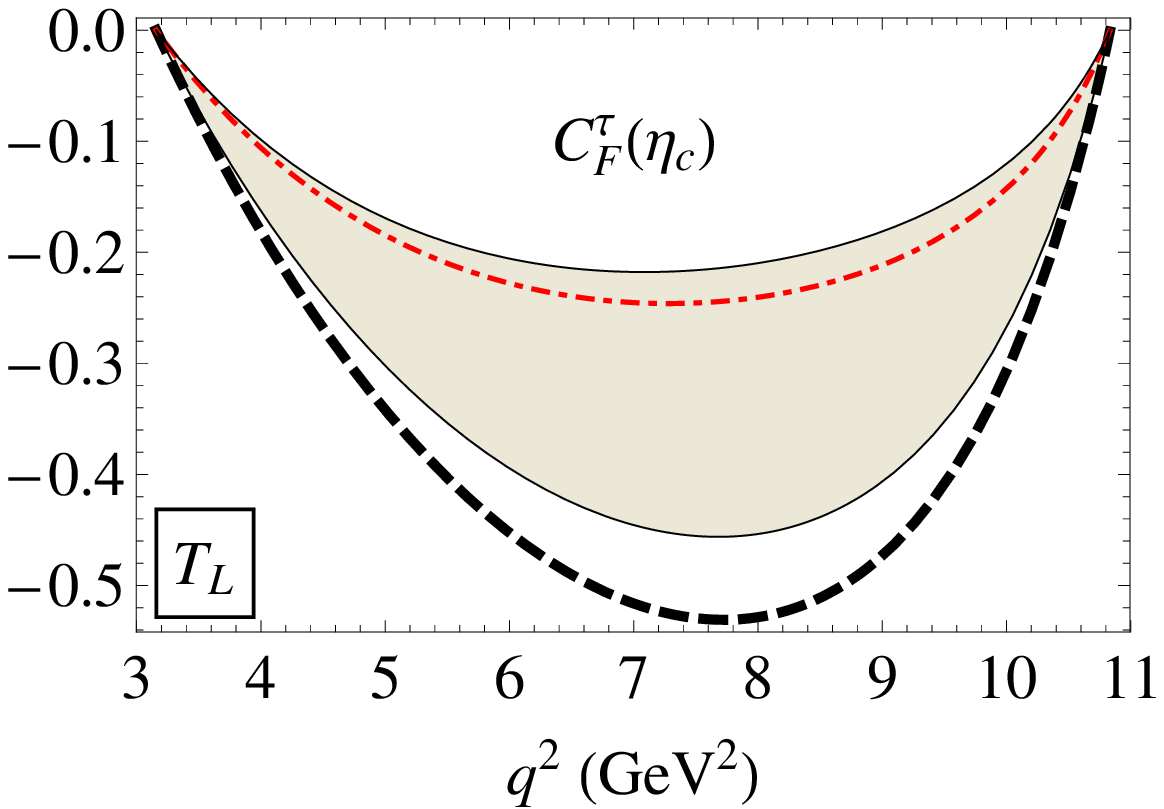}
&
\includegraphics[scale=0.4]{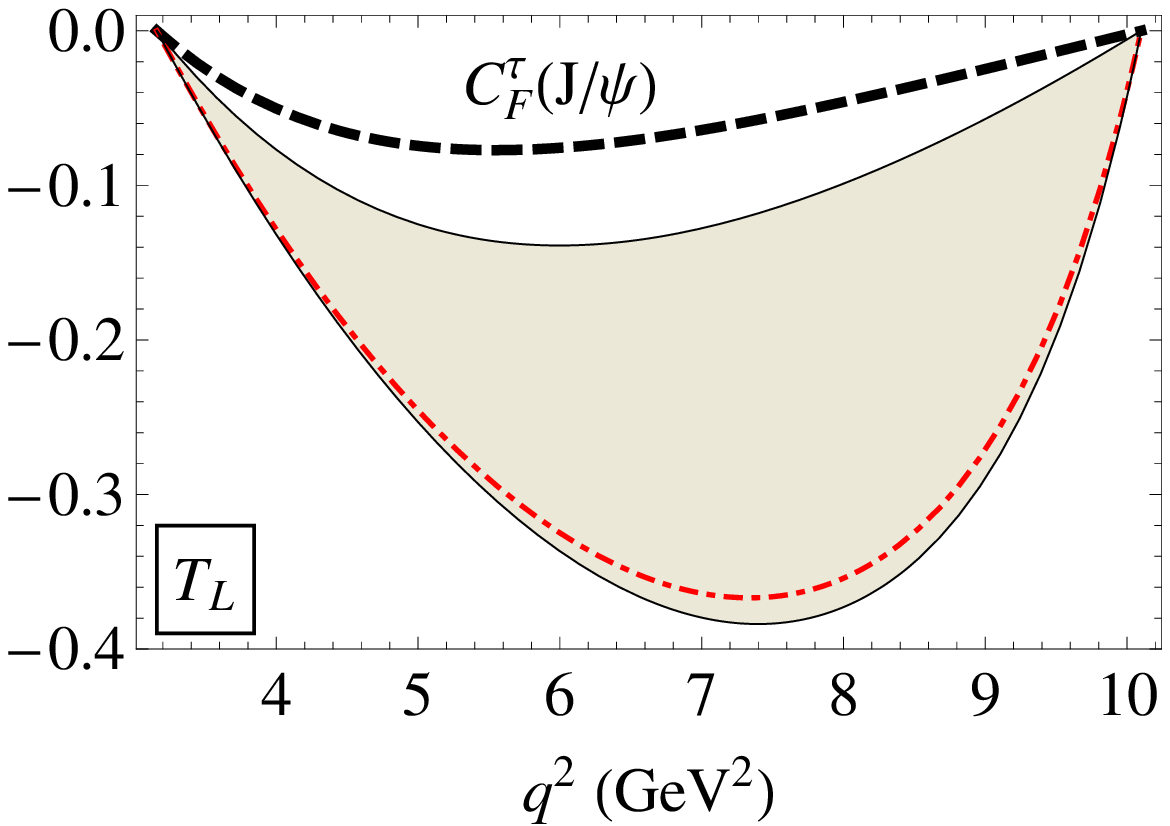}
&
\includegraphics[scale=0.4]{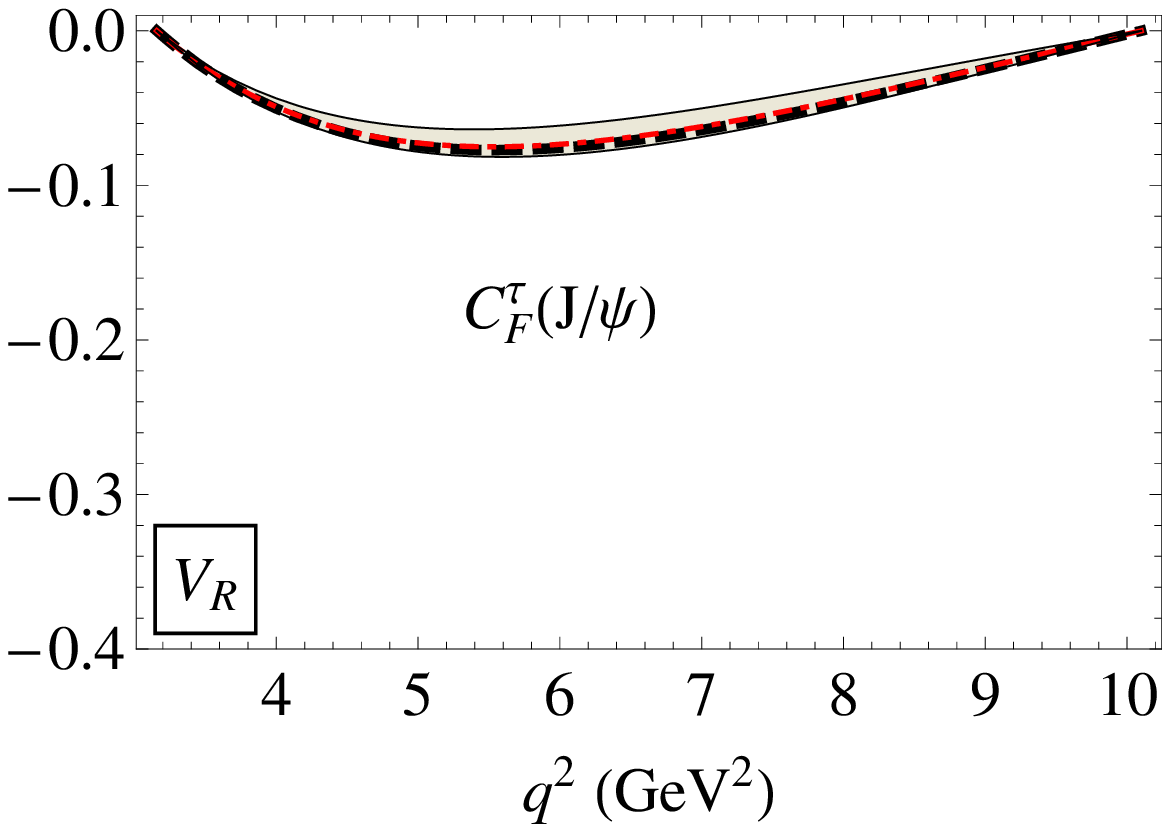}
\end{tabular}
\caption{Forward-backward asymmetry $\mathcal{A}_{FB}(q^2)$ (upper panels) and convexity parameter $C_F^\tau(q^2)$ (lower panels) for $B_c \to \eta_c\tau\nu$ and $B_c \to J/\psi\tau\nu$. Notations are the same as in Fig.~\ref{fig:R}. In the case of $B_c \to \eta_c\tau\nu$, $\mathcal{O}_{V_R}$ does not affect these observables.}
\label{fig:AFB-CFL}
\end{center}
\end{figure}

In the upper panels of Fig.~\ref{fig:AFB-CFL} we present the $q^2$ dependence of the forward-backward asymmetry
$\mathcal{A}_{FB}$. In the case of the $B_c\to J/\psi$ transition, the operator $\mathcal{O}_{V_R}$ tends to decrease $\mathcal{A}_{FB}$ and shift the zero-crossing point to greater values than the SM one, while the tensor operator $\mathcal{O}_{T_L}$ can enhance $\mathcal{A}_{FB}$ at high $q^2$. In the case $B_c\to \eta_c$, $\mathcal{O}_{V_R}$ does not affect $\mathcal{A}_{FB}$, while $\mathcal{O}_{T_L}$ tends to decrease $\mathcal{A}_{FB}$, especially at high $q^2$.

In the lower panels of Fig.~\ref{fig:AFB-CFL} we show the convexity parameter $C_F^\tau(q^2)$. It is seen that the operator $\mathcal{O}_{V_R}$ has a very small effect on $C_F^\tau$, and only in the case of $B_c\to J/\psi$. In contrast to this, $C_F^\tau$ is extremely sensitive to the tensor operator $\mathcal{O}_{T_L}$. In particular, $\mathcal{O}_{T_L}$ can change $C_F^\tau(J/\psi)$ by a factor of 4 at $q^2\approx 7.5~\text{GeV}^2$. Besides, $\mathcal{O}_{T_L}$ enhances the absolute value of $C_F^\tau(J/\psi)$, but reduces that of $C_F^\tau(\eta_c)$.

Similar to what has been discussed in Refs.~\cite{Chen:2017eby, Tanaka:2010se, Ivanov:2017mrj}, one can use the polarization of the $\tau$ in the semileptonic decays $B_c\to J/\psi(\eta_c)\tau\nu$ to probe for NP. The longitudinal ($L$), transverse ($T$), and normal ($N$) polarization components of the $\tau$ are defined as
\be
\label{eq:poldef}
P_i(q^2) = \frac{d\Gamma(s^\mu_i)/dq^2-d\Gamma(-s^\mu_i)/dq^2}{d\Gamma(s^\mu_i)/dq^2+d\Gamma(-s^\mu_i)/dq^2},\qquad i=L, N, T,
\en
where $s^\mu_i$ are the polarization four-vectors of the $\tau$ in the $W^-$ rest frame. One has
\be
s^\mu_L=\Big(\frac{|\vec{p}_\tau|}{m_\tau},\frac{E_\tau}{m_\tau}\frac{\vec{p}_\tau}{|\vec{p}_\tau|}\Big),\qquad
s^\mu_N=\Big(0,\frac{\vec{p}_\tau \times\vec{p}_2}{|\vec{p}_\tau \times\vec{p}_2|}\Big),\qquad 
s^\mu_T=\Big(0,\frac{\vec{p}_\tau \times\vec{p}_2}{|\vec{p}_\tau \times\vec{p}_2|}\times
\frac{\vec{p}_\tau}{|\vec{p}_\tau|}\Big).
\en
Here, $\vec{p}_\tau$ and $\vec{p}_2$ are the three-momenta of the $\tau$ and the final meson ($J/\psi$ or $\eta_c$), respectively, in the $W^-$ rest frame. A detailed analysis of the tau polarization with the help of its subsequent decays can be found in Refs.~\cite{Ivanov:2017mrj, Alonso:2016gym, Alonso:2017ktd}. 

\begin{figure}[htbp]
\begin{tabular}{ccc}
\includegraphics[scale=0.4]{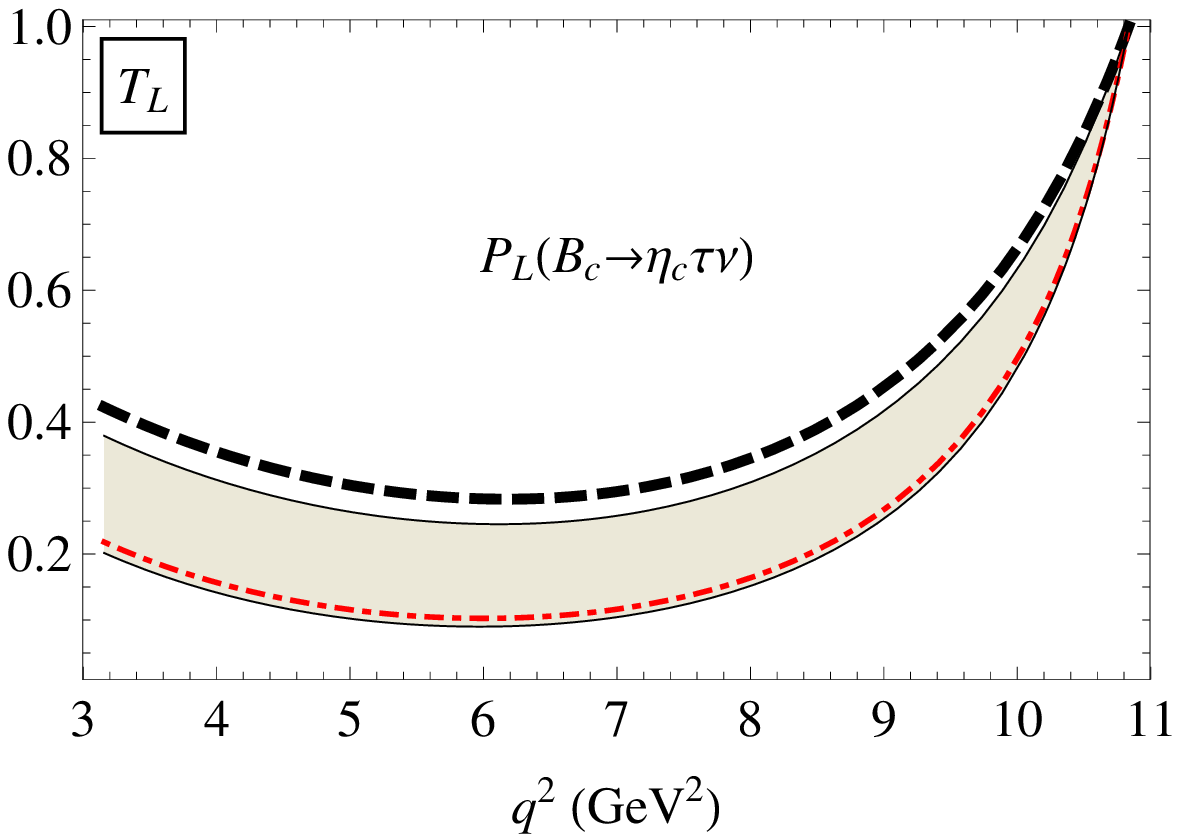}
& 
\includegraphics[scale=0.4]{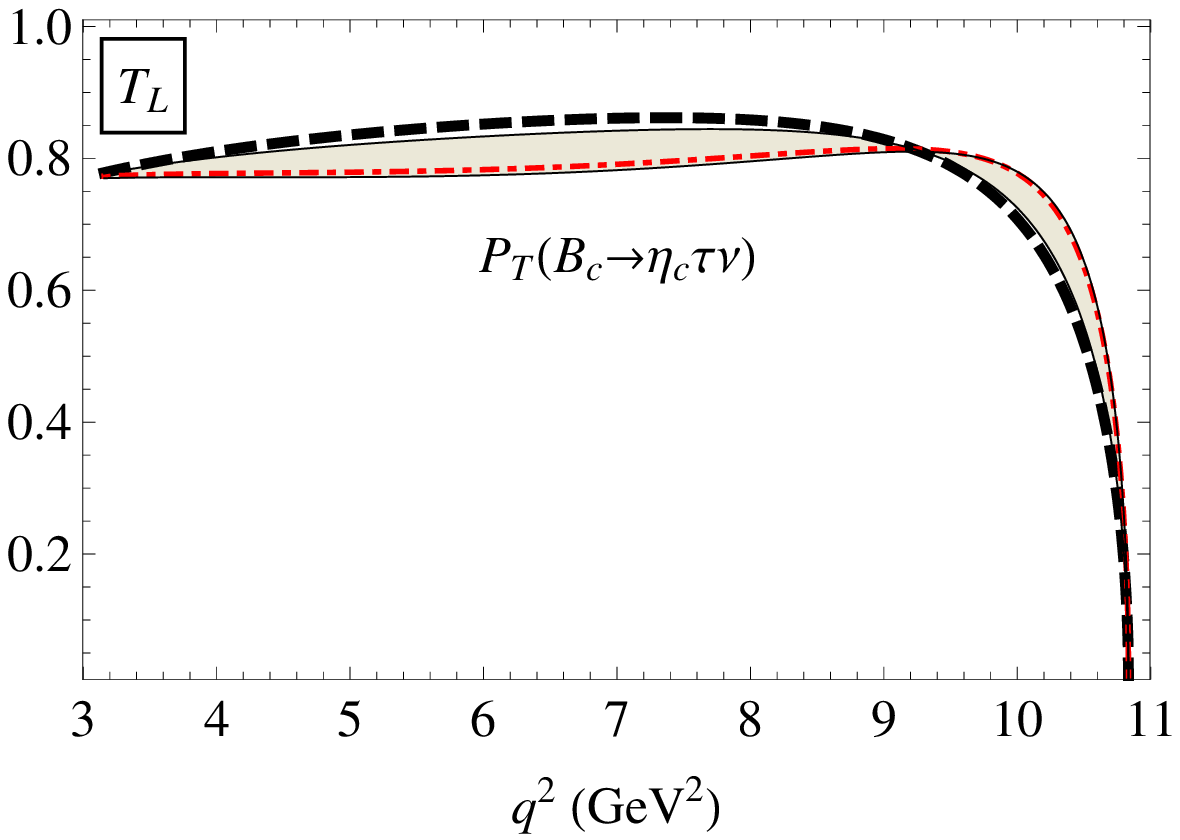}
& 
\includegraphics[scale=0.4]{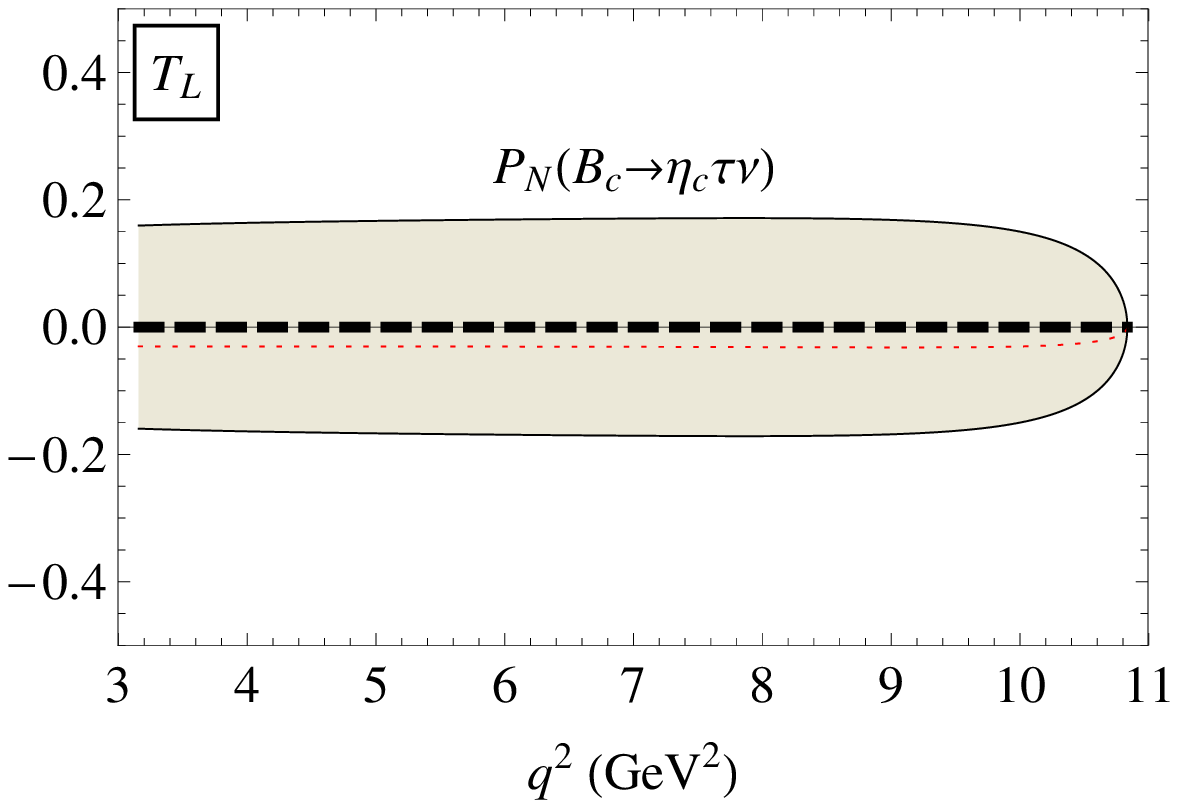}\\
\includegraphics[scale=0.4]{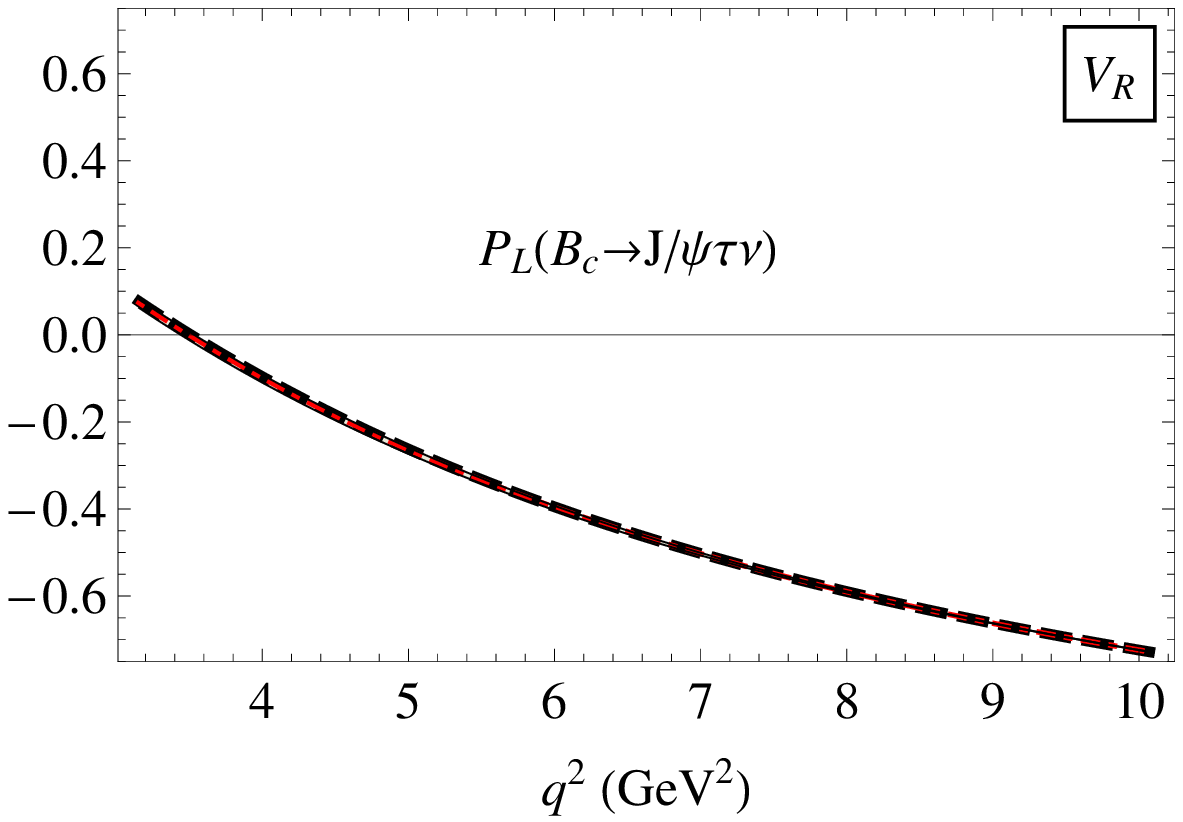}
& 
\includegraphics[scale=0.4]{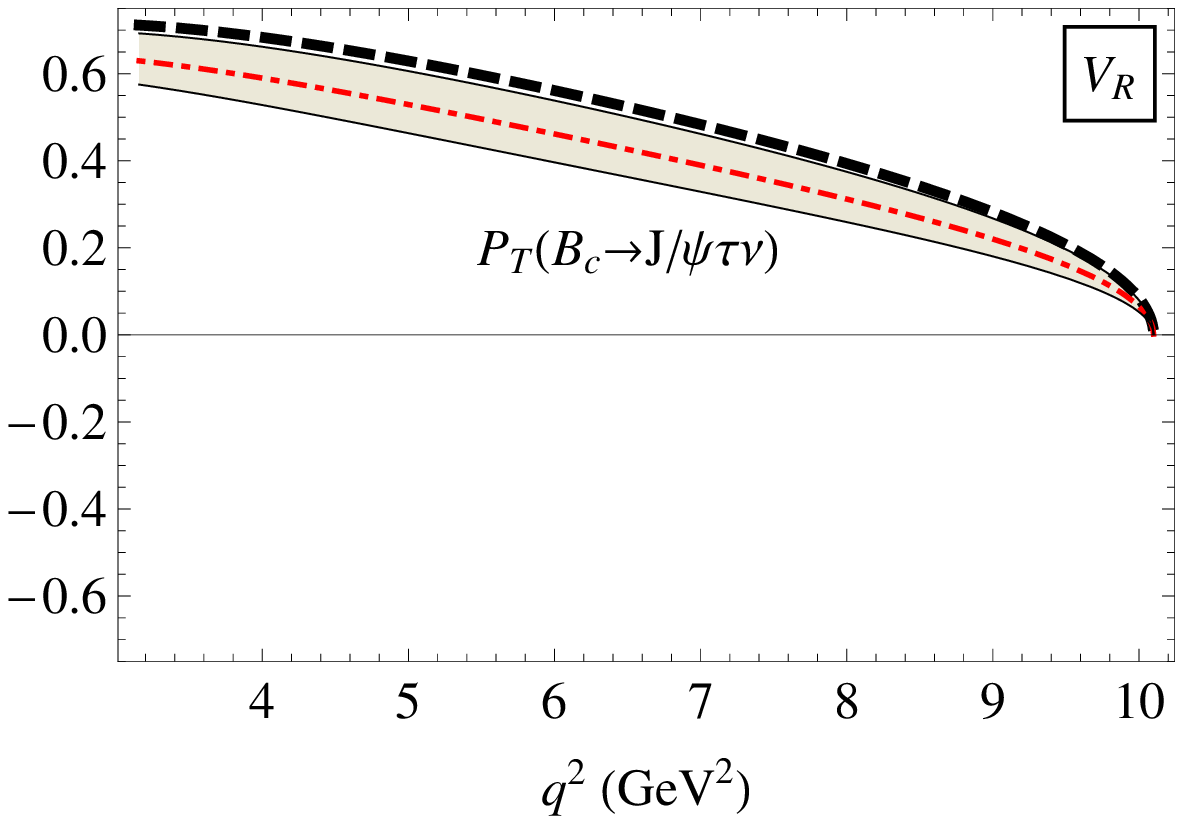}
&
\includegraphics[scale=0.4]{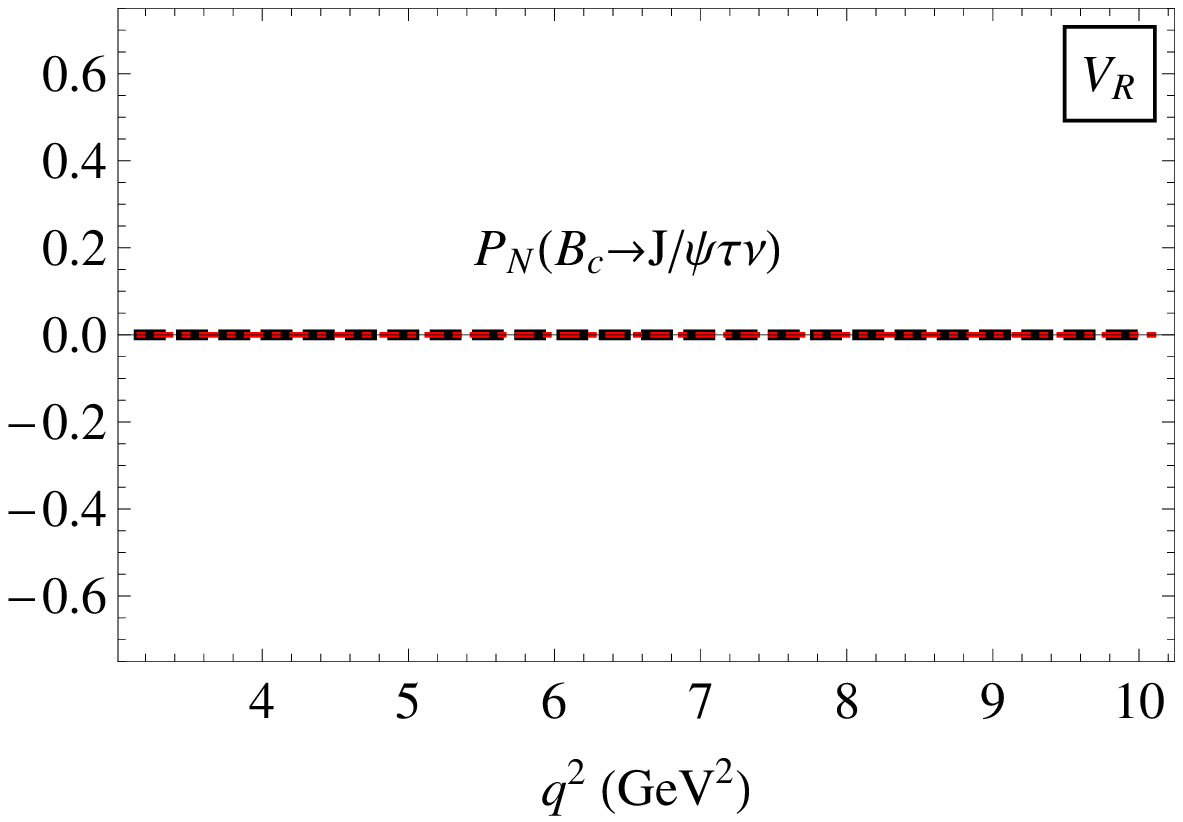}
\\
\includegraphics[scale=0.4]{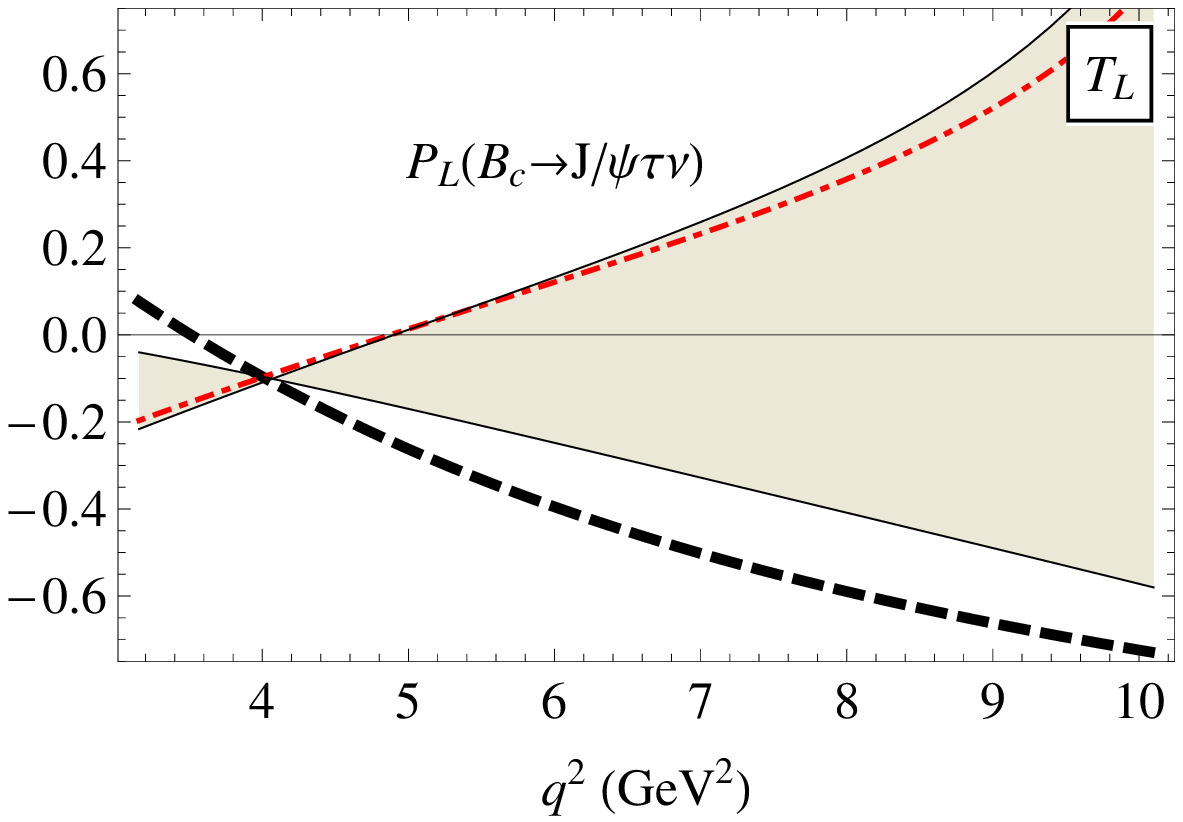}
& 
\includegraphics[scale=0.4]{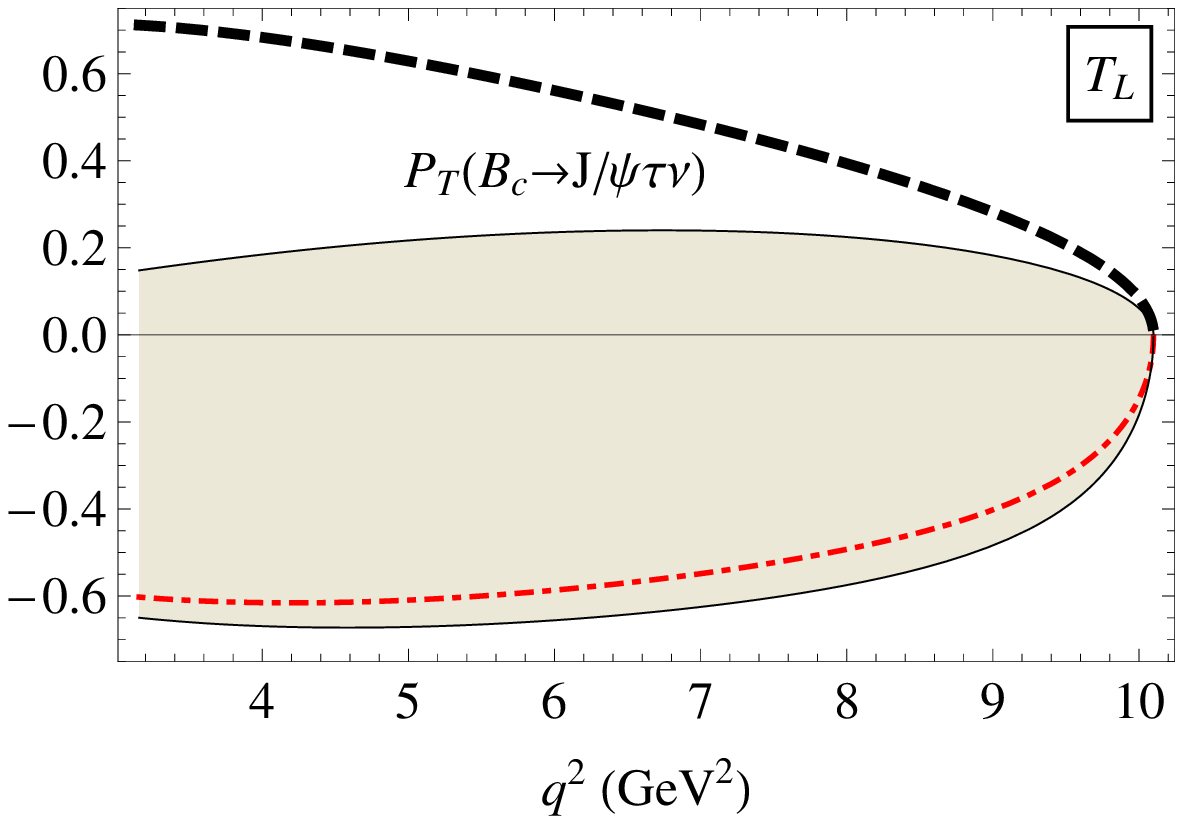}
& 
\includegraphics[scale=0.4]{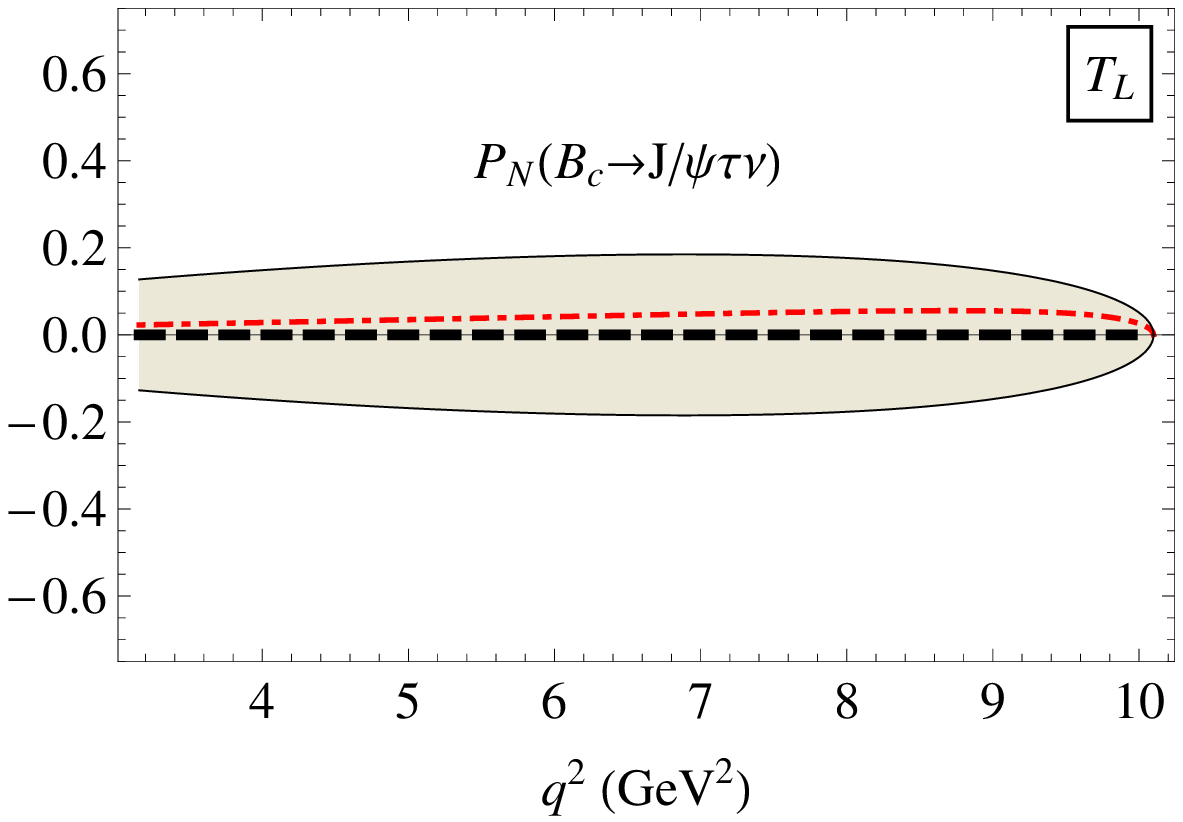}
\end{tabular}
\caption{Longitudinal (left), transverse (center), and normal (right) polarization of the $\tau$ in the decays $B_c \to \eta_c\tau\nu$ and $B_c \to J/\psi\tau\nu$. Notations are the same as in Fig.~\ref{fig:R}. In the case of $B_c \to \eta_c\tau\nu$, $\mathcal{O}_{V_R}$ does not affect these observables.}
\label{fig:pol}
\end{figure}

The $q^2$ dependence of the tau polarizations are presented in Fig.~\ref{fig:pol}. For easy comparison, the plots for each decay are scaled identically.  Several observations can be made here. First, the operator $\mathcal{O}_{V_R}$ affects only the tau transverse polarization in $B_c\to J/\psi\tau\nu$. Second, in both decays, all polarization components are very sensitive to the tensor operator $\mathcal{O}_{T_L}$. In the presence of $\mathcal{O}_{T_L}$, the longitudinal and transverse polarization of the tau in $B_c\to J/\psi\tau\nu$ can change their signs. And finally, the normal polarization, which is equal to zero in the SM, can become quite large when $\mathcal{O}_{T_L}$ is present.
The predictions for the mean polarization observables are summarized in Table~\ref{tab:pol-average} with the same notations as for Table~\ref{tab:ratio}.
\begin{table}[htbp] 
\caption{$q^{2}$ averages of the forward-backward asymmetry, the convexity parameter, the polarization components, and the total polarization.}
\begin{center}
\begin{tabular}{lccccccc}
\hline\hline
\multicolumn{7}{c}{ $B_c\to \eta_c$ }\\
\hline
& $<A_{FB}>$
&  $<C_{F}^\tau>$
&  $<P_{L}>$
&  $<P_{T}>$  
&  $<P_{N}>$   
&  $<|\vec P|>$
\\
\hline
SM \quad & $-0.36$  &$-0.43$
  & $0.36$ & $0.83$  & $0$& $0.92$
\\ 
$T_L$\quad
& $(-0.45,-0.37)$
& $(-0.38,-0.19)$
& $(0.16,0.32)$
& $(0.78,0.82)$
& $(-0.17,0.17)$
& $(0.81,0.90)$\\
\hline\hline
\multicolumn{7}{c}{ $B_c\to J/\psi$ }\\
\hline
&  $<A_{FB}>$ 
&  $<C_{F}^\tau>$  
&  $<P_{L}>$ 
&  $<P_{T}>$    
&  $<P_{N}>$  
& $<|\vec P|>$
\\
\hline
SM \quad& 0.03 & $-0.05$
 & $-0.51$& 0.43 &  0  & 0.70
\\
$V_R$\quad
& $(-0.09,0.01)$
& $(-0.05,-0.04)$
& $-0.51$
& $(0.30,0.41)$
& $0$
& $(0.62,0.69)$
\\
$T_L$\quad
& $(-0.10,0.01)$
& $(-0.31,-0.10)$
& $(-0.35,0.25)$
& $(-0.61,0.21)$
& $(-0.17,0.17)$
& $(0.23,0.70)$
\\
\hline\hline
\end{tabular}
\label{tab:pol-average}
\end{center}
\end{table}
\section{Summary and conclusions}
\label{sec:summary}
In the wake of recent measurements of the $B_c$ weak decays performed by the LHCb Collaboration, we have studied possible NP effects in the semileptonic decays $B_c\to J/\psi\tau\nu$ and $B_c\to \eta_c\tau\nu$ based on an effective Hamiltonian consisting of vector, scalar, and tensor four-fermion operators. The form factors parametrizing the corresponding hadronic transitions $B_c\to J/\psi$ and $B_c\to \eta_c$ have been calculated in the framework of the CCQM in the full kinematical region of momentum transfer. We have also provided a detailed comparison of our form factors with those of other authors and predicted the slope for the ratio of form factors $F_0(q^2)/F_+(q^2)$.  

Using the experimental data for the ratios $R_{D^{(\ast)}}$ and $R_{J/\psi}$ from the {\it BABAR}, Belle, and LHCb Collaborations, as well as the LEP1 result for the branching $\mathcal{B}(B_c\to\tau\nu)$, we have obtained the constraints on the Wilson coefficients characterizing the NP contributions.  It has turned out that at the level of $2\sigma$, the scalar coefficients $S_{L,R}$ are excluded, while the vector ($V_{L,R}$) and tensor ($T_L$) ones are still available. However, all coefficients are ruled out at $1\sigma$. It is worth mentioning that the constraints have been obtained under the assumption of one-operator dominance, where the interferences between different operators have been omitted.

Finally, within the $2\sigma$ allowed regions of the corresponding Wilson coefficients, we have analyzed the effects of the NP operators $\mathcal{O}_{V_L}$, $\mathcal{O}_{V_R}$, and $\mathcal{O}_{T_L}$ on various physical observables, namely, the ratios $R_{J/\psi}(q^2)$ and $R_{\eta_c}(q^2)$, the forward-backward asymmetry $\mathcal{A}_{FB}(q^2)$, the convexity parameter $C^\tau_F(q^2)$, and the polarizations of the $\tau$ in the final state. Some of the effects may help distinguish between NP operators. We have also provided predictions for the $q^2$ average of the mentioned observables, which will be useful for other theoretical studies and future experiments.


\begin{acknowledgments}
  The authors thank the Heisenberg-Landau Grant for providing support for their
  collaboration. M.A.I. acknowledges the financial support of the PRISMA Cluster of
  Excellence at the University of Mainz.  M.A.I. and C.T.T. greatly appreciate the warm
  hospitality of the Mainz Institute for Theoretical Physics (MITP) at the University of Mainz.
\end{acknowledgments}


\end{document}